\documentclass[longauth]{aa}
\usepackage{graphicx}
\usepackage[round]{natbib}
\usepackage[colorlinks=true,linkcolor=blue,urlcolor=blue,citecolor=red,pdftex]{hyperref}
\usepackage{threeparttable}
\usepackage[font={it},labelfont={bf}]{caption}
\usepackage[caption=false]{subfig}
\usepackage{times}
\usepackage{fixltx2e}

\title{The LOFAR radio environment}

\author{A.~R.~Offringa\inst{1,2}
  \and A.~G.~de~Bruyn\inst{1,3}
  \and S.~Zaroubi\inst{1}
  \and G.~van~Diepen\inst{3}
  \and O.~Martinez-Ruby\inst{1}
  \and P.~Labropoulos\inst{3}
  \and M.~A.~Brentjens\inst{3}
  \and B.~Ciardi\inst{4}
  \and S.~Daiboo\inst{1}
  \and G.~Harker\inst{5}
  \and V.~Jeli\'c\inst{3}
  \and S.~Kazemi\inst{1}
  \and L.~V.~E.~Koopmans\inst{1}
  \and G.~Mellema\inst{6}
  \and V.~N.~Pandey\inst{3}
  \and R.~F.~Pizzo\inst{3}
  \and J.~Schaye\inst{7}
  \and H.~Vedantham\inst{1}
  \and V.~Veligatla\inst{1}
  \and S.~J.~Wijnholds\inst{3}
  \and S.~Yatawatta\inst{3}
  \and P.~Zarka\inst{8}
  \and A.~Alexov\inst{9} % Amsterdam
  \and J.~Anderson\inst{10} % Bonn
  \and A.~Asgekar\inst{3} % ASTRON
  \and M.~Avruch\inst{11,1} % SRON + Kapteyn
  \and R.~Beck\inst{10} % Bonn
  \and M.~Bell\inst{12,17} % Sydney
  \and M.~R.~Bell\inst{4} % Garching
  \and M.~Bentum\inst{3} % Astron
  \and G.~Bernardi\inst{13,1} % Harvard
  \and P.~Best\inst{14} % Edinburgh
  \and L.~Birzan\inst{7} % Leiden
  \and A.~Bonafede\inst{15} % Bremen
  \and F.~Breitling\inst{16} % Potsdam
  \and J.~W.~Broderick\inst{17} % Southampton
  \and M.~Br\"uggen\inst{18,15} % Hamburg, Bremen
  \and H.~Butcher\inst{2,3} % Stromlo
  \and J.~Conway\inst{19} % Chalmers
  \and M.~de~Vos\inst{3} % Astron
  \and R.~J.~Dettmar\inst{20} % Ruhr
  \and J.~Eisloeffel\inst{21} % Tautenburg
  \and H.~Falcke\inst{22} % Radboud
  \and R.~Fender\inst{17} % Southampton
  \and W.~Frieswijk\inst{3} % Astron
  \and M.~Gerbers\inst{3} % Astron
  \and J.~M.~Griessmeier\inst{23,3} % CNRS
  \and A.~W.~Gunst\inst{3} % Astron
  \and T.~E.~Hassall\inst{17,24} % Southampton, Manchester
  \and G.~Heald\inst{3} % Astron
  \and J.~Hessels\inst{3} % Astron
  \and M.~Hoeft\inst{21} % Tautenburg
  \and A.~Horneffer\inst{10} % Bonn
  \and A.~Karastergiou\inst{25} % Oxford
  \and V.~Kondratiev\inst{3}
  \and Y.~Koopman\inst{3}
  \and M.~Kuniyoshi\inst{10} % Bonn
  \and G.~Kuper\inst{3}
  \and P.~Maat\inst{3}
  \and G.~Mann\inst{16} % Potsdam
  \and J.~McKean\inst{3}
  \and H.~Meulman\inst{3}
  \and M.~Mevius\inst{3}
  \and J.~D.~Mol\inst{3}
  \and R.~Nijboer\inst{3}
  \and J.~Noordam\inst{3}
  \and M.~Norden\inst{3} 
  \and H.~Paas\inst{3}
  \and M.~Pandey\inst{26,7} % Lyon + Leiden
  \and R.~Pizzo\inst{3}
  \and A.~Polatidis\inst{3}
  \and D.~Rafferty\inst{7} % Leiden
  \and S.~Rawlings\inst{26} % Oxford
  \and W.~Reich\inst{10} % Bonn
  \and H.~J.~A.~R\"ottgering\inst{7} % Leiden
  \and A.~P.~Schoenmakers\inst{3}
  \and J.~Sluman\inst{3}
  \and O.~Smirnov\inst{3,27} % RATT
  \and C.~Sobey\inst{10} % Bonn
  \and B.~Stappers\inst{24} % Manchester
  \and M.~Steinmetz\inst{16} % Potsdam
  \and J.~Swinbank\inst{9} % Amsterdam
  \and M.~Tagger\inst{23} % CNRS
  \and Y.~Tang\inst{3} % Astron
  \and C.~Tasse\inst{8} % Paris
  \and A.~van~Ardenne\inst{3,19} % Sweden
  \and W.~van~Cappellen\inst{3}
  \and A.~P.~van~Duin\inst{3}
  \and M.~van~Haarlem\inst{3}
  \and J.~van~Leeuwen\inst{3}
  \and R.~J.~van~Weeren\inst{7,3}
  \and R.~Vermeulen\inst{3}
  \and C.~Vocks\inst{16} % Postdam
  \and R.~A.~M.~J.~Wijers\inst{9} % Amsterdam
  \and M.~Wise\inst{3}
  \and O.~Wucknitz\inst{28,29,10} % Bonn2 + Bonn3 + Bonn
}

\institute{
  University of Groningen, Kapteyn Astronomical Institute, PO Box 800, 9700 AV Groningen, The Netherlands. \email{offringa@mso.anu.edu.au}
  \and Mount Stromlo Observatory, RSAA, Cotter Road, Weston Creek, ACT 2611, Australia. % 3
  \and ASTRON, PO Box 2, 7990 AA Dwingeloo, The Netherlands.
  \and Max-Planck Institute for Astrophysics, Karl-Schwarzschild-Strasse 1, 85748 Garching bei M\"unchen, Germany.
  \and Center for Astrophysics and Space Astronomy, University of Colorado, 389 UCB, Boulder, Colorado 80309-0389, USA.
  \and AlbaNova University Center, Department of Astronomy, SE-106 91 Stockholm, Sweden.
  \and Leiden Observatory, Leiden University, PO Box 9513, 2300 RA Leiden, The Netherlands.
  \and Observatoire de Paris, FR 92195 Meudon, France.
  \and University of Amsterdam, Astronomical Institute Anton Pannekoek, PO Box 94249, 1090 GE Amsterdam, The Netherlands.
  \and Max-Planck Institute for Astrophysics, P.O. Box 20 24, D-53010 Bonn, Germany.
  \and SRON Netherlands Institute for Space Research, PO Box 800, 9700 AV Groningen, The Netherlands.
  \and Sydney Institute for Astronomy, School of Physics A28, University of Sydney, NSW 2006, Australia.
  \and Harvard-Smithsonian Center for Astrophysics, 60 Garden Street, Cambridge, MA 02138 
  \and Royal Observatory Edinburgh, Blackford Hill, Edinburgh, EH9 3HJ, UK.
  \and Jacobs University Bremen, Campus Ring 1, 28759 Bremen, Germany.
  \and Astrophysical Institute Potzdam, An der Sternwarte 16, 14482 Potsdam, Germany.
  \and University of Southampton, University Road, Southampton SO17 1BJ, UK.
  \and University of Hamburg, Gojenbergsweg 112, 21029 Hamburg, Germany.
  \and Chalmers University of Technology, SE-412 96 Gothenburg, Sweden.
  \and Ruhr-Uuniversity Bochum, Universit\"atsstraße 150, 44801 Bochum, Germany.
  \and Th\"uringer Landessternwarte, Tautenburg Observatory, Sternwarte 5, D-07778 Tautenburg, Germany.
  \and Radboud University Nijmegen, Faculty of NWI, PO Box 9010, 6500 GL Nijmegen, The Netherlands.
  \and Centre national de la recherche scientifique, 3 rue Michel-Ange, 75794 Paris cedex 16, France.
  \and University of Manchester, Oxford Road, Manchester, M13 9PL, UK.
  \and University of Oxford, Wellington Square, Oxford OX1 2JD, UK.
  \and Centre de Recherche Astrophysique de Lyon, Observatoire de Lyon, 9 av
Charles Andr\'e, 69561 Saint Genis Laval Cedex, France.
  \and Rhodes University, RATT, Dep. Physics and Electronics, PO Box 94, Grahamstown 6140, South Africa.
  \and University of Bonn, Regina-Pacis-Weg 3, D-53012 Bonn, Germany.
  \and Argelander-Institut f\"ur Astronomie, Auf dem H\"ugel 71, 53121 Bonn, Germany.
} 

\newcommand{\degree}{\ensuremath{^\circ}}

\begin{document}

\abstract{}{This paper discusses the spectral occupancy for performing radio astronomy with the Low-Frequency Array (LOFAR), with a focus on imaging observations.}{We have analysed the radio-frequency interference (RFI) situation in two 24-h surveys with Dutch LOFAR stations, covering 30--78~MHz with low-band antennas and 115--163~MHz with high-band antennas. This is a subset of the full frequency range of LOFAR. The surveys have been observed with a 0.76~kHz~/~1~s resolution.}{We measured the RFI occupancy in the low and high frequency sets to be 1.8\% and 3.2\% respectively. These values are found to be representative values for the LOFAR radio environment. Between day and night, there is no significant difference in the radio environment. We find that lowering the current observational time and frequency resolutions of LOFAR results in a slight loss of flagging accuracy. At LOFAR's nominal resolution of 0.76~kHz and 1~s, the false-positives rate is about 0.5\%. This rate increases approximately linearly when decreasing the data frequency resolution.}{Currently, by using an automated RFI detection strategy, the LOFAR radio environment poses no perceivable problems for sensitive observing. It remains to be seen if this is still true for very deep observations that integrate over tens of nights, but the situation looks promising. Reasons for the low impact of RFI are the high spectral and time resolution of LOFAR; accurate detection methods; strong filters and high receiver linearity; and the proximity of the antennas to the ground. We discuss some strategies that can be used once low-level RFI starts to become apparent. It is important that the frequency range of LOFAR remains free of broadband interference, such as DAB stations and windmills.}


\maketitle

\section{Introduction}
\begin{figure*}
 \begin{center}
  \includegraphics[height=73mm]{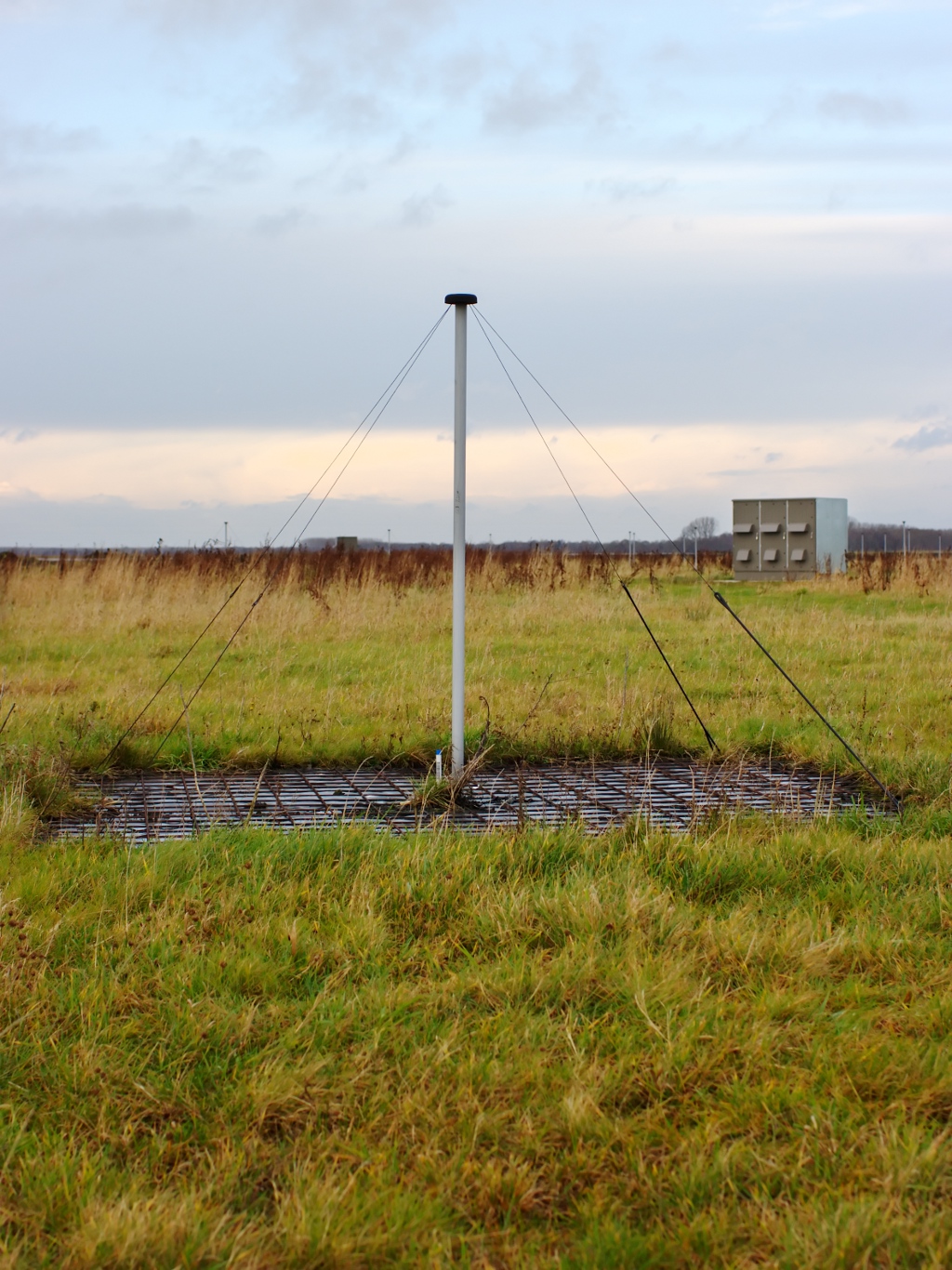}~%
  \includegraphics[height=73mm]{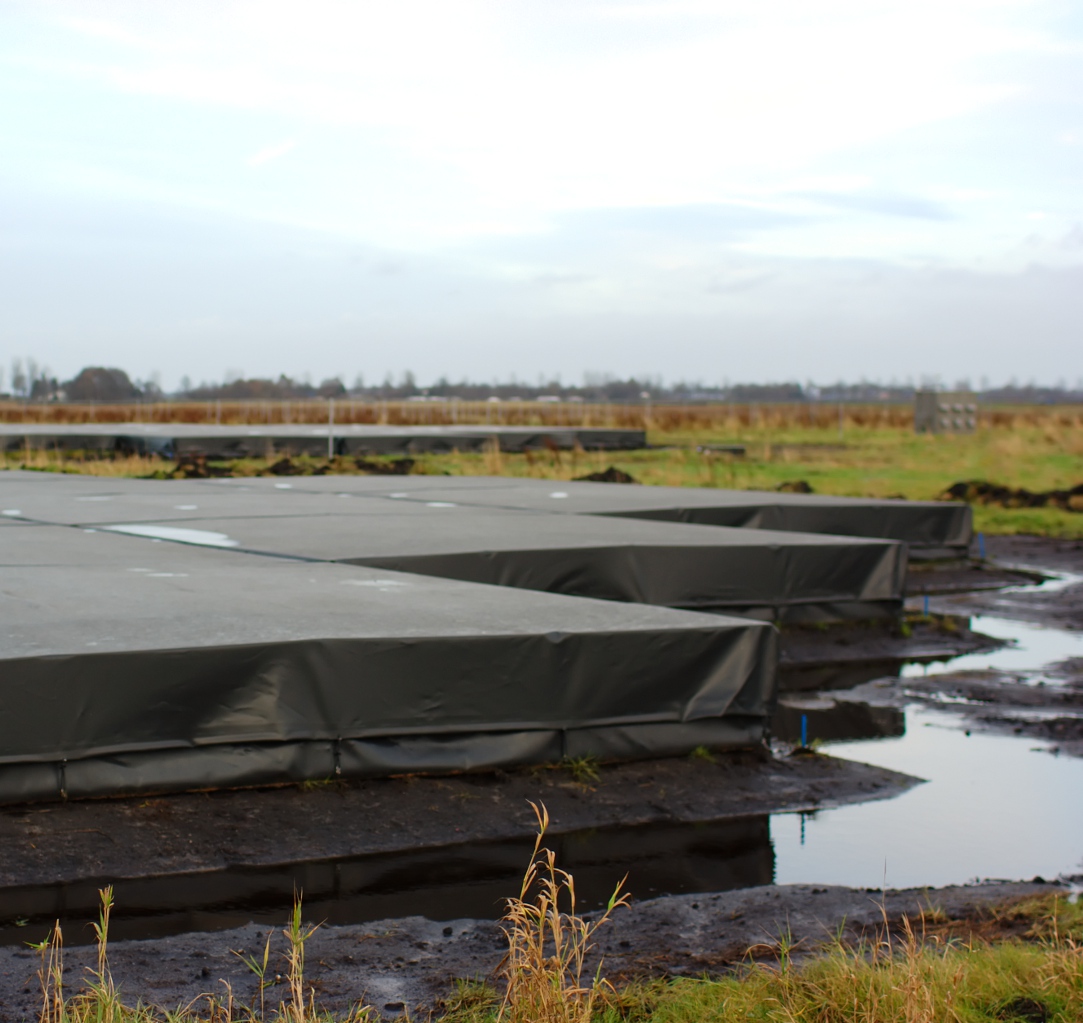}
 \end{center}
 \caption{Antenna types of the Low-Frequency Array. Left image: A low-band antenna with a cabin in the background. Right image: Part of a high-band antenna station, consisting of 24 tiles of 4 $\times$ 4 high-band antennas. }
 \label{fig:intro-lofar-foto}
\end{figure*}

The Low-Frequency Array (LOFAR) (van Haarlem et al., 2012, A\&A, in prep.) is a new antenna array that observes the sky from 10--80 and 110--240 MHz. It currently consists of 41 (validated) stations, while 7 more are planned. The number of stations are likely to increase further in the future. Of the validated stations, 33 stations are located in the Netherlands, 5 in Germany and one each in Sweden, the UK and France. A Dutch station consists of 96 dipole low-band antennas (LBA) that provide the 10--80 MHz range, and one or two fields totalling 48 tiles of 4x4 bow-tie high-band antennas (HBA) for the frequency range of 110-240 MHz. The two different antenna types are shown in Fig.~\ref{fig:intro-lofar-foto}. The international stations have an equal number of LBAs, but 96 HBA tiles. For the latest information about LOFAR, we refer the reader to the LOFAR website\footnote{The website of LOFAR is \url{http://www.lofar.org/} .}.

\begin{figure}%
 \begin{center}\includegraphics[height=95mm]{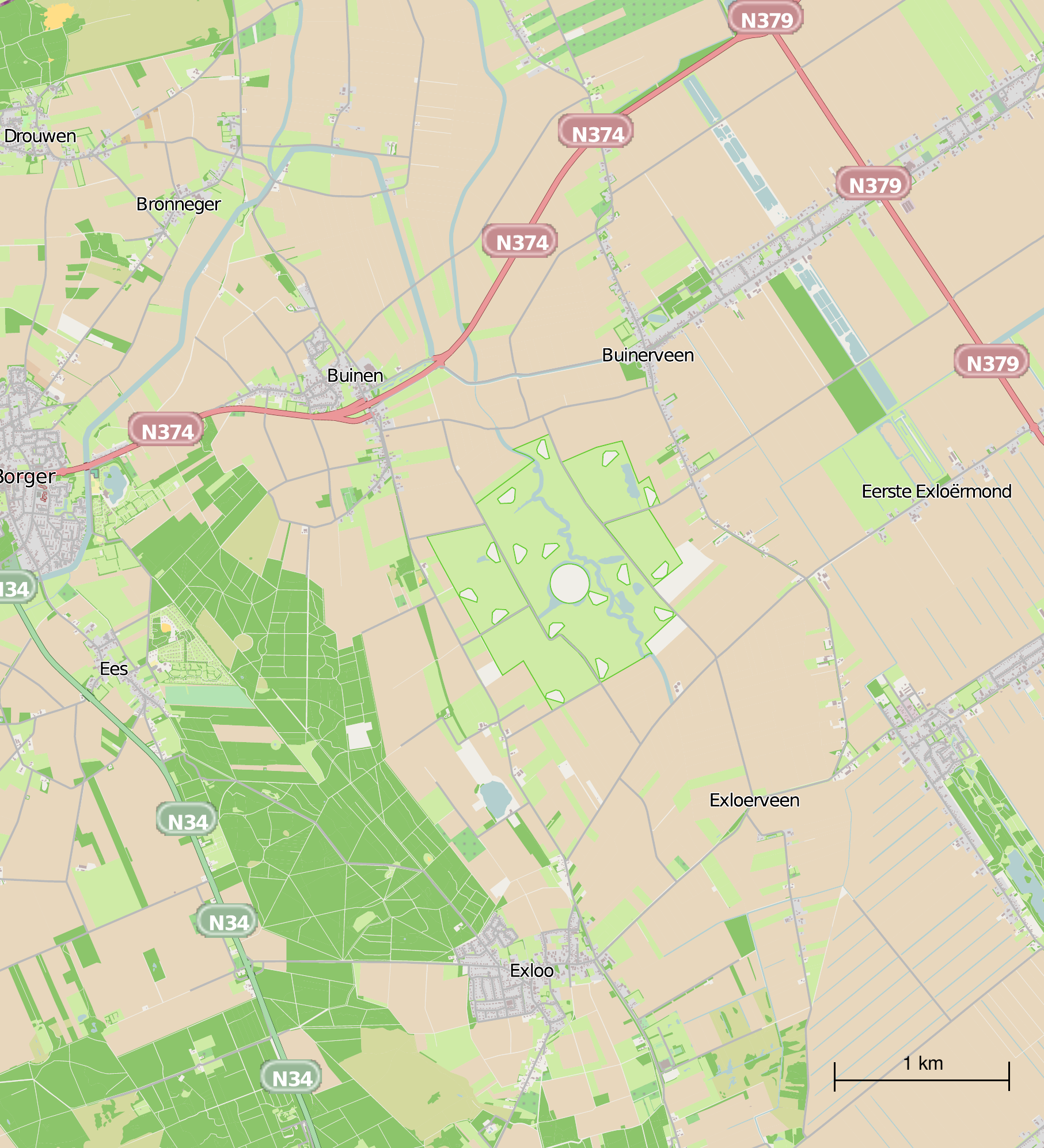}\end{center}%
 \caption{Map of the LOFAR core and its surroundings. The circular peninsula in the centre is the Superterp. Several other stations (triangular footprints) are visible as well. (source: OpenStreetMap) }%
 \label{fig:env-lofar-core-map}%
\end{figure}

The core area of LOFAR is located near the village of Exloo in the Netherlands, where the station density is at its highest. The six most densely packed stations are on the Superterp, an elevated area surrounded by water. It is an artificial island of about 350~m in diameter that is situated about 3~km North of Exloo. A map of LOFAR's surroundings is given in Fig.~\ref{fig:env-lofar-core-map}. Exloo is a village in the municipality of Borger-Odoorn in the province of Drenthe. Drenthe is mostly a rural area, and is sparsely populated relative to the rest of the Netherlands, with an average density of 183~persons/km$^2$ over 2,680~km$^2$ in 2011\footnote{From the website of the province of Drenthe,\\\url{http://www.provincie.drenthe.nl/} .}. Nevertheless, the radio-quiet zone of 2~km around the Superterp is relatively small, and households exists within 1~km of the Superterp. The distance from households to the other stations is even smaller in certain instances. Therefore, contamination of the radio environment by man-made electromagnetic radiation has been a major concern for LOFAR \citep{bregman-lofar-concept-design, bentum-lofar-rfi-mitigation}. Because this radiation interferes with the celestial signal of interest, it is referred to as radio-frequency interference (RFI). Such radiation can originate from equipment that radiates deliberately, such as citizens' band (CB) radio devices and digital video or audio broadcasting (DVB or DAB), but can also be due to unintentionally radiating devices such as cars, electrical fences, power lines and wind turbines \citep{assessment-of-LOFAR-RFI}.

During the hardware design phase of LOFAR, careful consideration was given to ensure that the signal would be dominated by the sky noise \citep{dipole-array-sensitivity-cappellen,sky-noise-limited-wijnholds}. This included placing shielding cabinets around equipment on site to minimise self-interference; making sure that RFI would not drive the amplifiers and analogue-digital converters (ADCs) into the non-linear regime; applying steep analogue filters to suppress the FM bands and frequencies below 10~MHz; and applying strong digital sub-band filters to localise RFI in frequency. Optionally, an additional analogue filter can be turned on to filter frequencies below 30~MHz.

Numerous techniques have been suggested to perform the task of RFI excision. They include using spatial information provided in interferometers or multi-feed systems to null directions \citep{multichannel-rfi-mitigation, ellingson-spatial-nulling-2002, hampson-spatial-nulling-2002, boonstra-dissertation, spatial-filtering-parkes-multibeam}; removing the RFI by using reference antennas \citep{adaptive-cancellation}; and blanking out unlikely high values at high time resolutions \citep{chi-square-time-blanking-weber, multichannel-rfi-mitigation, wsrt-rfims, pulse-blanking}. During post-processing, RFI excision can consist of detecting the RFI in time, frequency and antenna space, and ignoring the contaminated data in further data processing. This step is often referred to as ``data flagging''. Because of the major increase in resolution and bandwidth of observatories, leading to observations of tens of terabytes, manual data flagging is no longer feasible. Automated RFI flagging pipelines can solve this problem \citep{effelsberg-rfi-mitigation, LOFAR-RFI-pipeline}. Alternative RFI strategies might be required for the detection of transients \citep{exoplanet-detection-with-rfi, spatial-filtering-parkes-multibeam-for-pulses}.

Now that LOFAR deployment is nearly complete, commissioning observations have started and preliminary results show that the choice of LOFAR's site has not seriously degraded the data quality. For example, both the LOFAR-EoR project \citep{de-bruyn-eor-ursi-2011} and the LOFAR project on pulsars and fast transients \citep{lofar-pulsars} report that the data quality, in terms of the achieved sensitivity and calibratability, is as expected. Moreover, new algorithms and a pipeline have been implemented to automatically detect RFI with a high accuracy \citep{post-correlation-rfi-classification,LOFAR-RFI-pipeline}. Preliminary results have shown that by using these algorithms, only a few percent of the data is lost due to RFI \citep{LOFAR-RFI-pipeline}.

In this article, we study two 24-h RFI surveys: one for the 30--78~MHz low-band regime and one for the 115--163~MHz high-band regime. The observations were carried out in standard imaging mode in which visibilities are integrated to a time resolution of a second and have a spectral resolution of 0.76~kHz. In Sect.~\ref{sec:env-lofar}, we start by describing the relevant technical details of the LOFAR observatory. In Sect.~\ref{sec:spectrum-management}, a brief analysis of the spectrum allocation situation relevant for LOFAR is presented. In Sect.~\ref{sec:env-processing}, we describe the methods that are used to process and analyse the two data sets. Sect.~\ref{sec:env-survey-data} describes the details of the RFI observations that are used in this article. In Sect.~\ref{sec:env-results} we describe the observational results of the two RFI surveys. We also compare them with other observations to assess whether they are representative in Sect.~\ref{sec:env-compare}. In Sect.~\ref{sec:env-discussion}, we discuss the results and draw conclusions about the LOFAR RFI environment.

\section{LOFAR} \label{sec:env-lofar}
In this section, we will briefly describe the design details of LOFAR that are relevant for the impact of RFI. For further technical details, we refer the reader to \citet{lofar-system-design} and van Haarlem et al. (2012, A\&A, in prep.).

LOFAR consists of stations of clustered LBA and HBAs. The signals from the dual
polarisation LBAs are amplified with low-noise amplifiers (LNA), and are
subsequently transported over a coax cable to the electronics cabinet. The
signals from the HBAs are amplified and processed by an analogue beamformer,
which forms the beams for a tile of four times-four dipoles, before being sent
to the cabinet. In the cabinet the signal from either the LBAs or the HBAs is
band-pass filtered, digitised with a 12-bit ADC and one or more station beams
are formed.

Before station beams are formed, the HBA or LBA signals are split into 512~sub-bands of 195~kHz of bandwidth, of which 244 can be selected for further processing. Other modes can optionally be processed through different signal paths. The sub-bands are formed by using a poly-phase filter (PPF) that is implemented inside the station cabinet by using field-programmable gate arrays (FPGAs). This allows for very flexible observing configurations \citep{blue-gene-romein}. The 244~sub-band signals are transported over a dedicated wide-area network (WAN) to a Blue Gene/P (BG/P) supercomputer located in the city of Groningen. Currently, the samples are sent as 16 bit integers. However, because the transfer rate is limited to about 3 Gbit/s, the transport limits the total observed bandwidth to 48~MHz. Eight-bit and four-bit modes are scheduled to be implemented in late 2012, which would allow the transfer of 96-MHz and 192-MHz of bandwidth respectively. Multiple beams can be used, in which case the sum of the bandwidth over all beams is limited by these values.

The BG/P supercomputer applies a second PPF that increases the frequency resolution typically by a factor of 256, yielding a spectral resolution of 0.76~kHz. During this stage, the first of the 256 channels is lost for each sub-band, due to the way the PPF is implemented. Next, the BG/P supercomputer correlates each pair of stations, integrates the signal over time and applies a preliminary pass-band correction \citep{bandpass-correction-lofar-romein}, which corrects for the response of the first (station level) poly-phase filter. Finally, the correlation coefficients are written to the discs of the LOFAR Central Processing II (CEP2) cluster.

The partitioning into sub-bands is used to distribute data over the hard discs of the computing nodes on the CEP2 cluster. For storage of observations in imaging mode, LOFAR uses the CASA\footnote{CASA is the Common Astronomy Software Applications package, developed by an international consortium of scientists under the guidance of NRAO. Website: \url{http://casa.nrao.edu/}} measurement set (MS) format. The first step of post-processing of the observations is RFI excision. This is performed by the AOFlagger pipeline that is described in~\S\ref{sec:lofar-detection-strategy}. Further processing, such as averaging, calibration and imaging, ignores RFI contaminated data.

\section{Spectrum management} \label{sec:spectrum-management}
In the Netherlands, the radio spectrum use is regulated by the governmental agency ``Agentschap Telecom'', that falls under the Dutch Ministry of Economic Affairs, Agriculture and Innovation. This body maintains the registry of the Dutch spectrum users, which can be obtained from their website.\footnote{The website of the Agentschap Telecom from which the spectrum registry can be obtained is\\ \url{http://www.agentschaptelecom.nl/}.}

The other countries that participate in the International LOFAR Telescope have similar bodies, and the Electronic Communications Committee\footnote{The website of the Electronic Communications Committee, which registers spectrum usage at the European level, is\\\url{http://www.cept.org/ecc}, office: \url{http://www.ero.dk/}.} (ECC), a component of the European Conference of Postal and Telecommunications Administrations (CEPT), registers the use of the spectrum at the European level. Most of the strong and harmful transmitters are allocated in fixed bands for all European countries, such as the FM radio bands, satellite communication, weather radars and air traffic communication. However, even though the allocations of the countries are similar, the usage of the allocated bands can differ. For example, several 1.792~MHz wide channels between 174 and 195~MHz are registered as terrestrial digital audio broadcasting (T-DAB) bands by the ECC. These frequencies are correspondingly allocated to T-DAB both in the Netherlands and in Germany. However, these bands are currently used in Germany, but not yet in the Netherlands. Nevertheless, the range of 216--230~MHz is actively used for T-DAB in the Netherlands. This range corresponds with T-DAB bands 11A--11D and 12A--12D, each of which is 1.792~MHz wide. These transmitters are extremely harmful for radio astronomy. Because they are wideband and have a 100\% duty cycle and band usage, they do not permit radio observations. Digital video broadcasts (DVB) are similar, but occupy bands between 482 and 834~MHz (UHF channels 21--66). They are therefore outside the observing frequency range of LOFAR. Other transmitters are intermittent or occupy a narrow bandwidth, and therefore do allow radio-astronomical observations.

\begin{table}
\caption{Short list of allocated frequencies in the Netherlands in the range 10--250 MHz (source: Agentschap Telecom)}\label{tbl:radio-allocated-services}
\begin{center}
\begin{tabular}{ll}
\hline
\hline
\textbf{Service type} & \textbf{Frequency range(s) in MHz} \\
\hline
Time signal & 10, 15, 20 \\
Air traffic & 10--22, 118--137, 138--144 \\
Short-wave radio broadcasting & 11--26 \\
Military, maritime, mobile & 12--26, 27--61, 68--88, 138--179\\
Amateur & 14, 50--52, 144--146 \\
CB radio & 27-28 \\
Modelling control & 27--30, 35, 40--41 \\
Microphones & 36--38, 173--175 \\
\textbf{Radio astronomy} & 13, 26, 38, 150--153 \\
Baby monitor (portophone) & 39--40 \\
Broadcasting & 61--88 \\
Emergency & 74, 169--170 \\
Air navigation & 75, 108--118 \\
FM radio & 87--108 \\
Satellites & 137--138, 148--150 \\
Navigation & 150 \\
Remote control & 154 \\
T-DAB & 174--230 \\
Intercom & 202--209 \\
\hline
\hline
\end{tabular}
\end{center}
\end{table}

A short list of services with their corresponding frequencies is given in Table~\ref{tbl:radio-allocated-services}. Only a few small ranges are protected for radio-astronomy. The lowest ranges are 13.36--13.41, 25.55--25.67 and 37.5--38.25~MHz. These bands are useful for observing the Solar corona and Jovian magnetosphere, although they are too narrow, as the Sun and Jupiter emit broadband spectra. At higher LOFAR frequencies, the 150--153~MHz band is available for radio astronomy. Although the 10--200~MHz bandwidth is mostly allocated to other services, many of these --- such as baby monitors --- are used for short distance communication, and are therefore of low power. In addition, services such as the Citizens' Band (CB) radio transmitters have a low duty cycle (especially during the night) and individual transmissions are of limited bandwidth. The most problematic services for radio astronomy are therefore the FM radio (87.5--108~MHz), T-DAB (174--230~MHz) and the emergency pager (169.475--169.4875 and 169.5875--169.6~MHz) services. The FM radio range is excised by analogue filters. The emergency pager was found to be the strongest source in the spectrum. Therefore, the LOFAR signal path was designed to be able to digitise its signals correctly, i.e., without introducing non-linearities.

Around the LOFAR core, a radio-quiet zone has been established that is enforced by the province of Drenthe. The area is split into two zones. The inner zone of 2 km diameter around the core enforces full radio quietness. A ``negotiation zone'' with a diameter of about 10 km around the core requires negotiation before transmitters can be placed.\footnote{The radio quiet zones are marked on ``Kaart 12 --- overige aanduidingen'' of the environment plan of Drenthe.}

\section{Processing strategy} \label{sec:env-processing}
Processing an observation and acquiring an overview of the radio environment requires RFI detection statistics and quality assessment of the remaining data. In the following subsection, we address the detection strategy and the tools that we use for the detection. This is followed by a description of the methods for statistical analysis of RFI and data.

\subsection{Detection strategy} \label{sec:lofar-detection-strategy}
For RFI detection, LOFAR uses the AOFlagger pipeline. This pipeline iteratively estimates the contribution of the sky by using a Gaussian high-pass filter in the time-frequency domain of a single baseline. Subsequently, the {\tt Sum\-Thresh\-old} method \citep{post-correlation-rfi-classification} is used to detect line-shaped features in the same domain. A morphological operation named the scale-invariant rank (SIR) operator \citep{scale-invariant-rank-operator} is used to extent the flags into neighbouring regions that are also likely to be affected. The 4 cross-correlations (XX, XY, YX, YY) from the differently-polarised feeds are flagged individually. Finally, if a sample is flagged in one of the cross-correlations, it is also flagged in the corresponding other cross-correlations.

The pipeline is developed in the context of the LOFAR Epoch of Reionisation key science project and was described with more detail in \citet{LOFAR-RFI-pipeline}. Compared to the strategy described there, several optimisations were made to increase the speed of the flagger. One of the changes was to use a more stable and faster algorithm to compute the morphological SIR operator \citep{scale-invariant-rank-operator}. Another change was to implement several algorithms using the ``streaming single-instruction-multiple-data extensions'' (SSE) instruction set extension. The combined optimisations led to a decrease in the computational requirements of approximately a factor of 3, and the pipeline is input-output (IO) limited. To decrease the IO overhead, the pipeline was embedded in the ``New default pre-processing pipeline'' (NDPPP) \footnote{See \S 5 of ``The {LOFAR} Imaging Cookbook: Manual data reduction with the imaging pipeline'', ed. R.~F.~Pizzo et al., 2012, Astron technical document.}, which performs several tasks, such as data averaging and checking data integrity.

\begin{figure*}
\begin{center}
\includegraphics[width=18cm]{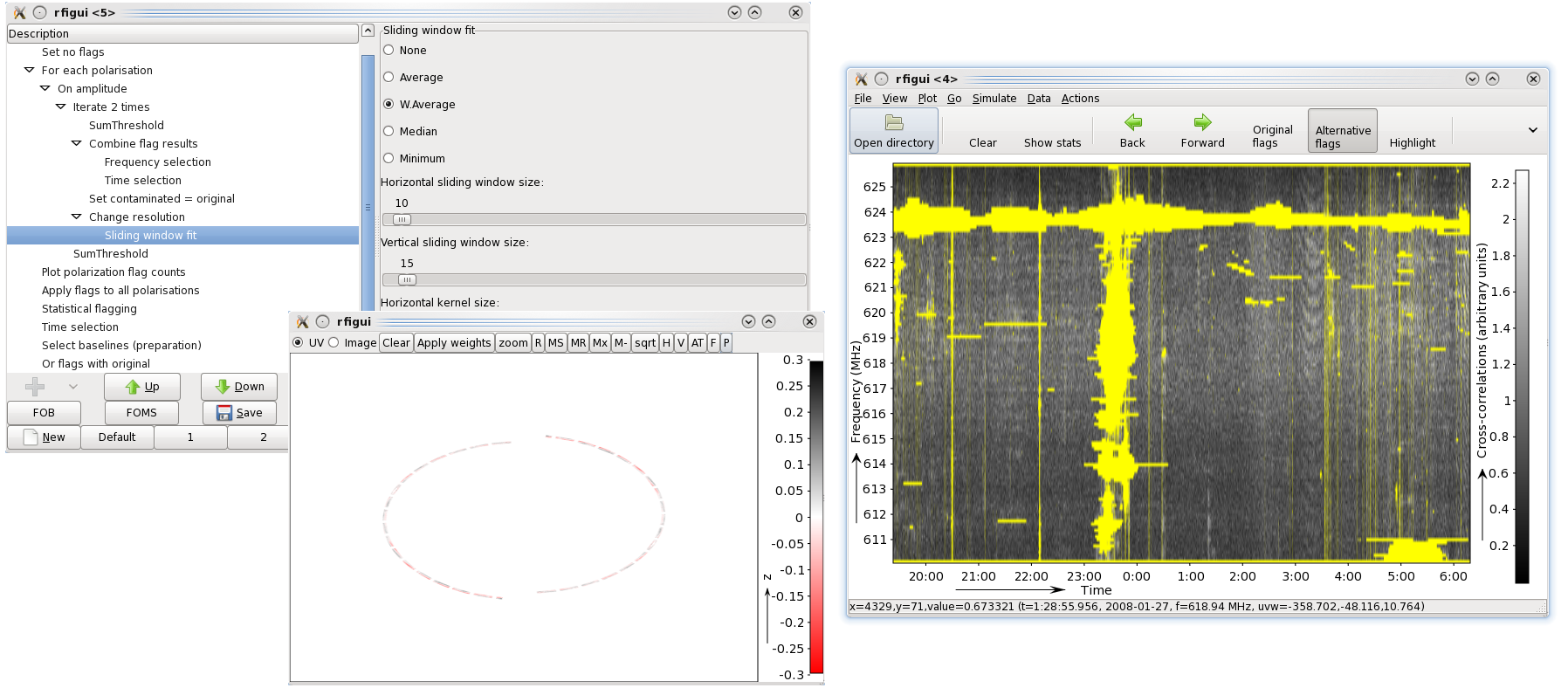}
\caption{Example snapshot of \texttt{rfigui}, which can be used to optimise the pipeline steps and tuning parameters. On the right is the main window showing the spectrum and flags (in yellow) of the selected baseline --- in this case a GMRT data set. The left bottom window shows the $uv$~track that this baseline covers. The upper-left window depicts the script with the actions that are performed, which can be edited interactively.}
\label{fig:aorfigui-example}
\end{center}
\end{figure*}

The AOFlagger package\footnote{The AOFlagger package is distributed under the GNU General Public License version 3.0, and can be downloaded from \url{http://www.astro.rug.nl/rfi-software}.} consists of three parts: (i)~the library that implements the detection pipeline and allows for its integration into pipelines of other observatories and NDPPP; (ii)~a stand-alone executable that runs the standard pipeline or a customised version; and (iii)~a graphical user interface (GUI) that can be used to analyse the flagging results on a baseline-by-baseline basis and optimise the various parameters of the pipeline (see Fig.~\ref{fig:aorfigui-example}). The GUI was used extensively to optimise the accuracy of the pipeline. It has also been used for implementing customised strategies for data from other observatories. This has for example led to successful flagging of data from the Westerbork Synthesis Radio Telescope (WSRT) \citep{post-correlation-rfi-classification} and the Giant Metrewave Radio Telescope (GMRT) (A.~D.~Biggs, personal communication, Sept. 2011). Similar application of the AOFlagger on single dish data from the Parkes radio telescope also shows good initial results (J.~Delhaize, personal communication, Aug. 2012).

For the data processing in this paper, we have used the original full resolution sets and applied the stand-alone flagger.

\subsection{RFI and quality statistics}
Assessing the quality of observations that have a volume of tens of terabyte is a non-trivial task. For example, simple operations such as calculating the mean or the root mean square (RMS) of the data are IO limited. Although these tasks can be distributed over multiple nodes if available, accessing all data of an observation still takes on the order of a few hours for large observations.

A generic solution was designed to assess the RFI situation and quality of an observation, by combining RFI statistics with other system statistics in a single platform. It consists of the following three parts: (1) a standardised storage format for the statistics; (2) software to collect the statistics; and (3) software to interpret the statistics. We will briefly describe each of these.
\begin{enumerate}
\item \textbf{The standardised storage format}: this was implemented as a format description of the so-called ``quality tables'' extension to the measurement set format\footnote{described by Offringa in the technical report ``Proposal for adding statistics sub-tables to a measurement set'', University of Groningen, 2011}. The CASA measurement set format allows adding custom tables, and we used this feature to add the statistics to the set. These statistics can be retrieved quickly without having to read the main data.

\begin{figure}
\begin{center}
\includegraphics[width=75mm]{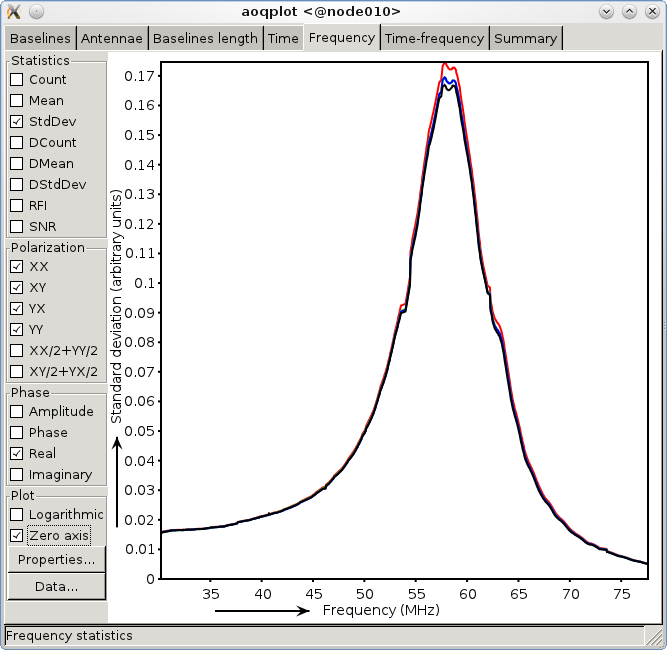}
\caption{The \texttt{aoqplot} tool displays the statistics interactively. In this case it shows the visibility standard deviation over frequency for a LBA observation.}
\label{fig:aoqplot}
\end{center}
\end{figure}

The quality tables contain statistics as a function of frequency, time, baseline index and polarisation. The stored values allow calculation of the fraction of detected RFI in the data (RFI occupancy), the mean (signal strength), the standard deviation and the differential standard deviation as a function of time, frequency, baseline index and polarisation. The mean and standard deviation are calculated for the RFI-free samples. The differential standard deviation describes the standard deviation of the noise by subtracting adjacent channels. Since the uncorrelated channels are only 0.76~kHz wide, the difference between adjacent channels should contain no significant contribution of the celestial signal, and are therefore a measure of the celestial and receiver noise (times $\sqrt{2}$).

\item \textbf{Software to collect the statistics}: We have implemented software that collects the statistics and writes them in the described format to the measurement set. A statistics collector was added to the NDPPP averaging step. Since NDPPP is performed on most LOFAR imaging observations, all observations will thereafter have these quality tables. NDPPP is slowed down by a few per cent because the statistics have to be calculated, which is acceptable. A stand-alone tool (``\texttt{aoquality}'') is available in the AOFlagger package that can collect the statistics without having to run NDPPP.

\item \textbf{Software to interpret the statistics}: Once the statistics are in the described format in the tables, tools are required to read and display the quality tables. Inside the AOFlagger package is an executable (``\texttt{aoqplot}'') that performs this task: it takes either a single measurement set or an observation file that specifies where the measurement sets are located, and opens a window in which various plots can be shown and the selection can be interactively changed. An example of the plotting tool is shown in Fig.~\ref{fig:aoqplot}.
\end{enumerate}

\section{Description of survey data} \label{sec:env-survey-data}
Table~\ref{table:rfi-survey-specs} lists the specifications of the two 24-h RFI surveys. The number of stations used in the HBA observation was reduced to limit the volume of data. More stations were included in the LBA observation. The sets were observed at standard LOFAR time and frequency resolutions of 1~s and 0.76~kHz respectively. In both sets, the observed field was the North Celestial Pole (NCP). This field does not have a bright radio source and it is therefore easier to detect the RFI due to the absence of strong rapidly oscillating visibility fringes. Therefore, it is to be noted that if an observation is affected by very strong off-axis sources, the level of false positives might by higher than reported in this article. Only in a very few observations we see effects of strong sources that impact flagging accuracy, and this can be solved by using a customised version of the AOFlagger. The NCP field does not require tracking and fringe stopping. This might also affect the detected occupancy, since some RFI might be averaged out when applying fringe stopping. Finally, the NCP field is a good field to observe with LOFAR, because it is always at a reasonably high elevation and it is also one of the targets of the LOFAR Epoch of Reionisation project (Yatawatta et al., in prep.).

\begin{table}
\caption{Survey data set specifications}\label{table:rfi-survey-specs}
\begin{center}
\begin{tabular}{lrr}
                 & \textbf{LBA set}& \textbf{HBA set} \\
\hline
\hline
Observation date & 2011-10-09      & 2010-12-27 \\
Start time       & 06:50 UTC       & 0:00 UTC \\
Length           & 24 h            & 24 h \\
Time resolution  & 1 s             & 1 s \\
\hline
Frequency range  &  30.1--77.5 MHz & 115.0--163.3 MHz\\
Frequency resolution & 0.76 kHz    & 0.76 kHz \\
Number of stations &  33           & 14 \\
\hspace{5mm} \footnotesize{\textit{Core}}   & \footnotesize{24} & \footnotesize{8} \\
\hspace{5mm} \footnotesize{\textit{Remote}} & \footnotesize{9} & \footnotesize{6} \\
\hline
Total size       & 96.3 TB        & 18.6 TB \\
Field            & NCP             & NCP \\
\hline
\hline
\end{tabular}
\end{center}
\end{table}

\begin{figure}
\begin{center}
\noindent\includegraphics[width=85mm]{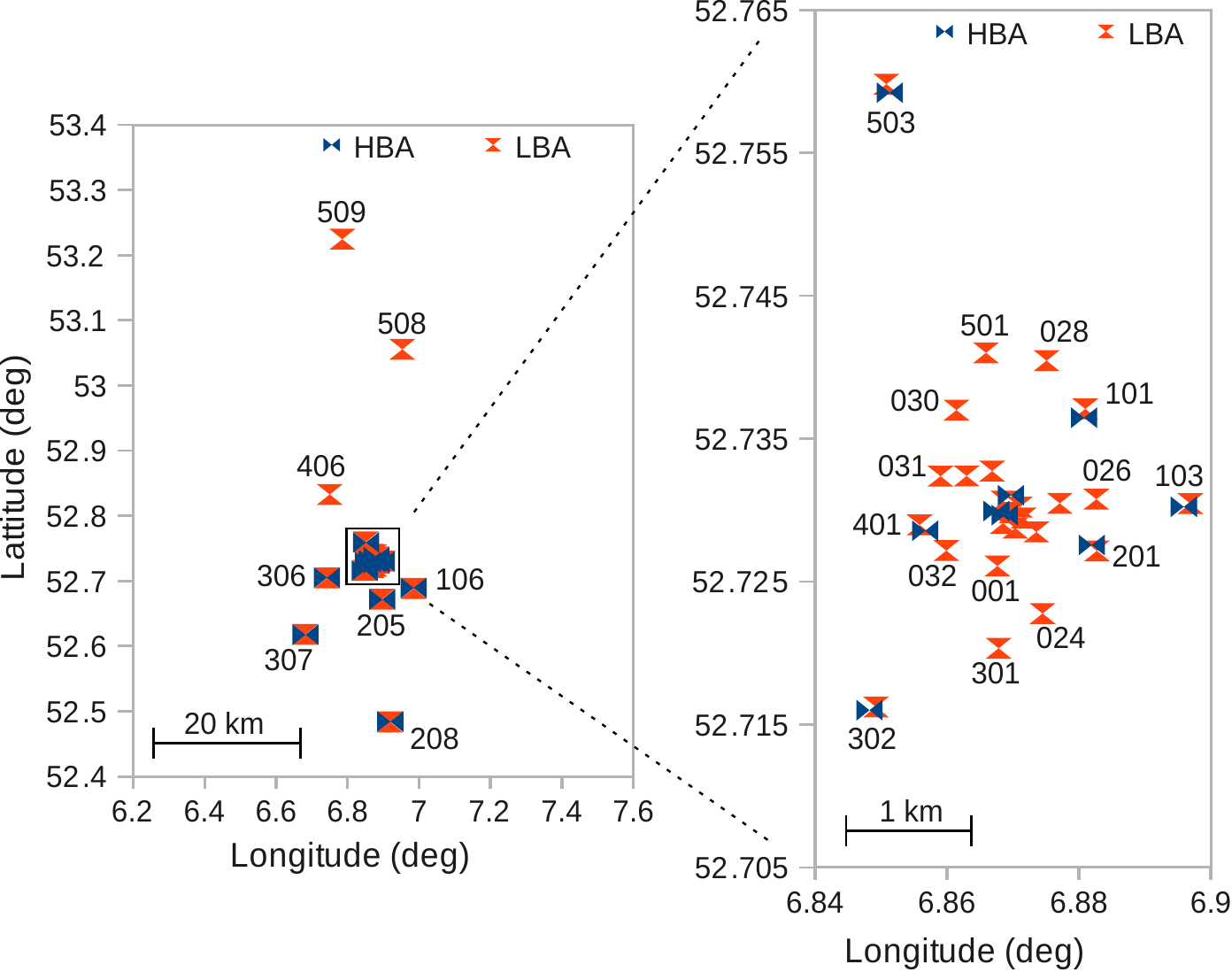}
\caption{Overview of the geometric distribution of the stations used for the RFI survey. Numbers next to the station symbols denote the station numbers.}
\label{fig:survey-antennaemap}
\end{center}
\end{figure}

Fig.~\ref{fig:survey-antennaemap} shows the locations of the stations that have been used in the two surveys. For the HBA set, the stations were selected to make sure that various baseline lengths were covered and the stations had a representative geometrical coverage. Due to the inclusion of additional core stations in the LBA set, the LBA set includes more short baselines.

In the LBA set, 6 sub-bands were corrupted due to two nodes on the LOFAR CEP2 cluster that failed during observing, causing six gaps of approximately 0.2~MHz in the 48-MHz frequency span of the observation. It is expected that such losses will be less common in future observations.

\section{Results} \label{sec:env-results}
In this section, we discuss the achieved performance of the flagger, look at the RFI implications of the surveys individually and analyse their common results.

\subsection{Performance}
We have used the LOFAR Epoch of Reionization (EoR) cluster (see Labropoulos et al., in prep.) to perform the data analysis. This cluster consists of 80 nodes with two hyperthreaded quad-core 2.27-GHz CPUs, two NVIDIA Tesla C1060 GPU's, 12 GB memory per node and 2 or 3 discs of approximately 2 terabyte (TB) each. The cluster is optimised for computationally intensive (GPU) tasks, such as advanced calibration and data inversion. Because it has relatively slow discs that are not in a redundant configuration (such as RAID), the cluster is not ideal for flagging, as flagging is computationally conservative and dominated by IO. To make sure the flagging would not interfere with computational tasks that were running on the cluster at that time, we chose to use only 3 CPU cores out of the 16 available cores, thus a fraction of 3/16 of the entire CPU power of the cluster. Flagging the 96-TB observation with version 2.0.1 of the AOFlagger took 40 hours, of which 32 hours were spend on reordering the observation, which consists only of reading and writing to the hard discs, and the remaining 8 hours were spent on actual flagging. 

\subsection{LBA survey} \label{sec:lba-survey}
\begin{figure*}
\begin{center}
\includegraphics[height=75mm]{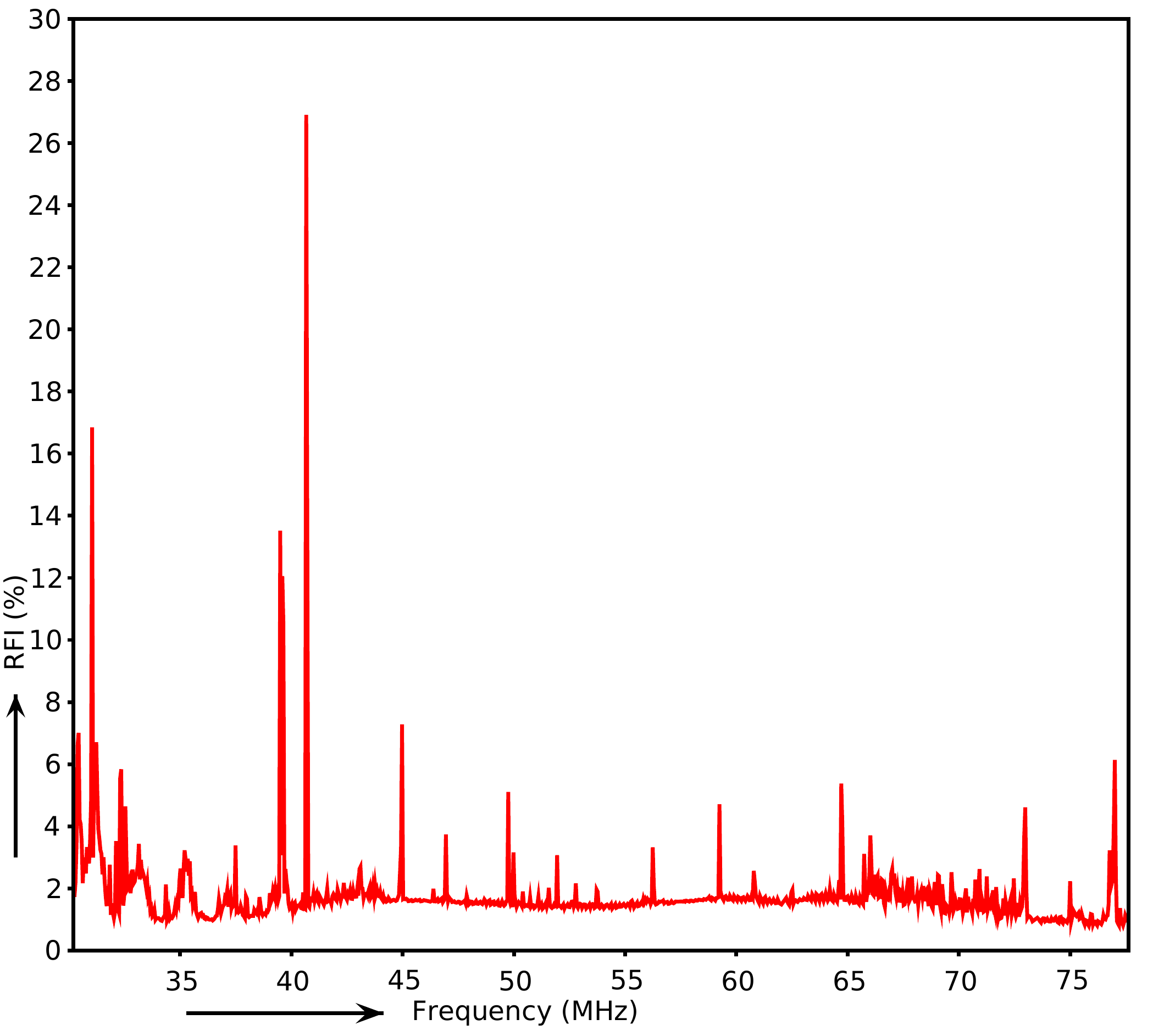}\hspace{0.3pc}
\includegraphics[height=75mm]{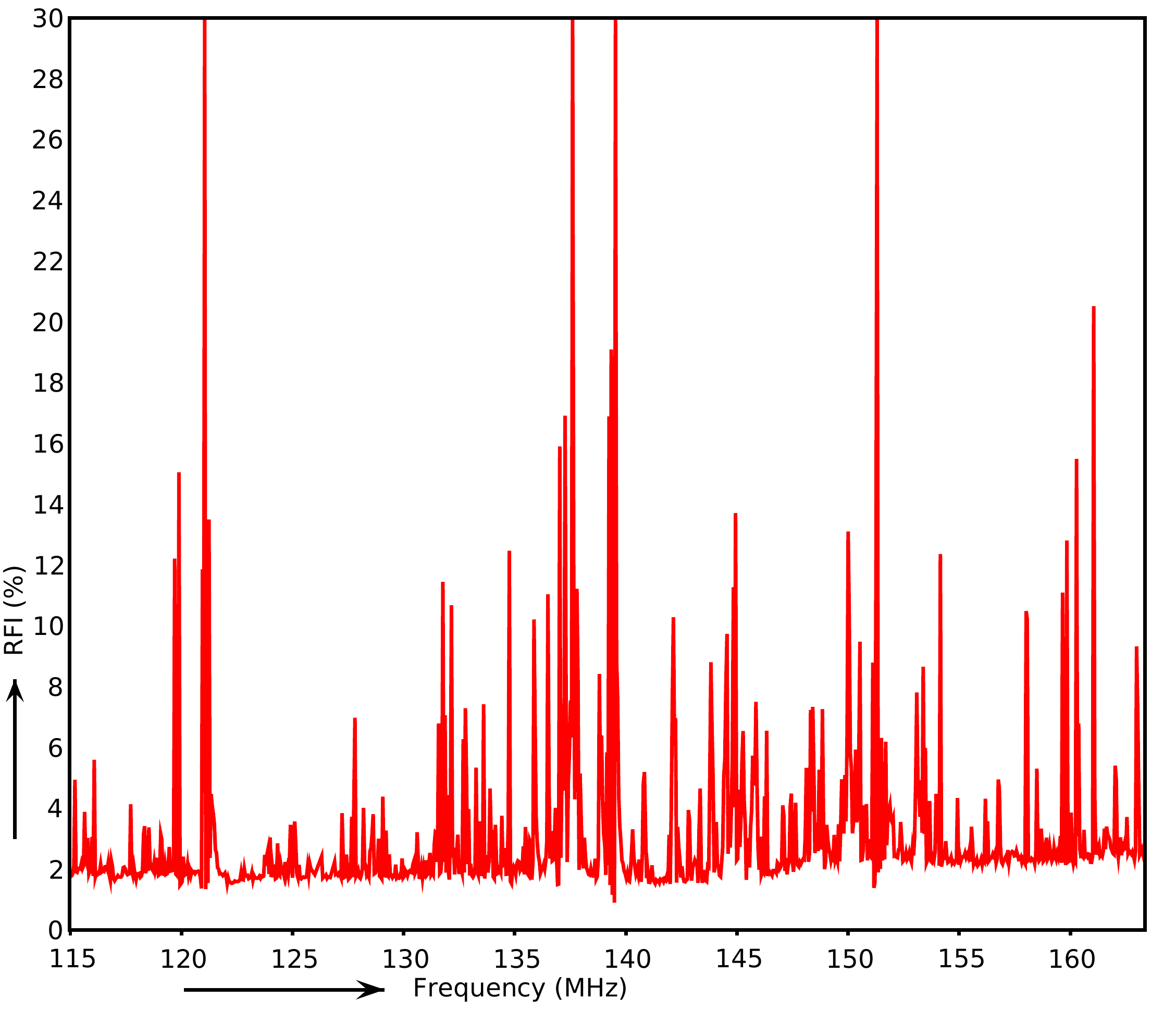}
\caption{The detected RFI occupancy spectra for both RFI surveys. Each data sample in the plot contains 48 kHz of data.}
\label{fig:survey-rfi-freq-plot}
\end{center}
\end{figure*}

\begin{figure*}
\begin{center}
\noindent\hspace*{-0.5cm}\includegraphics[width=36pc]{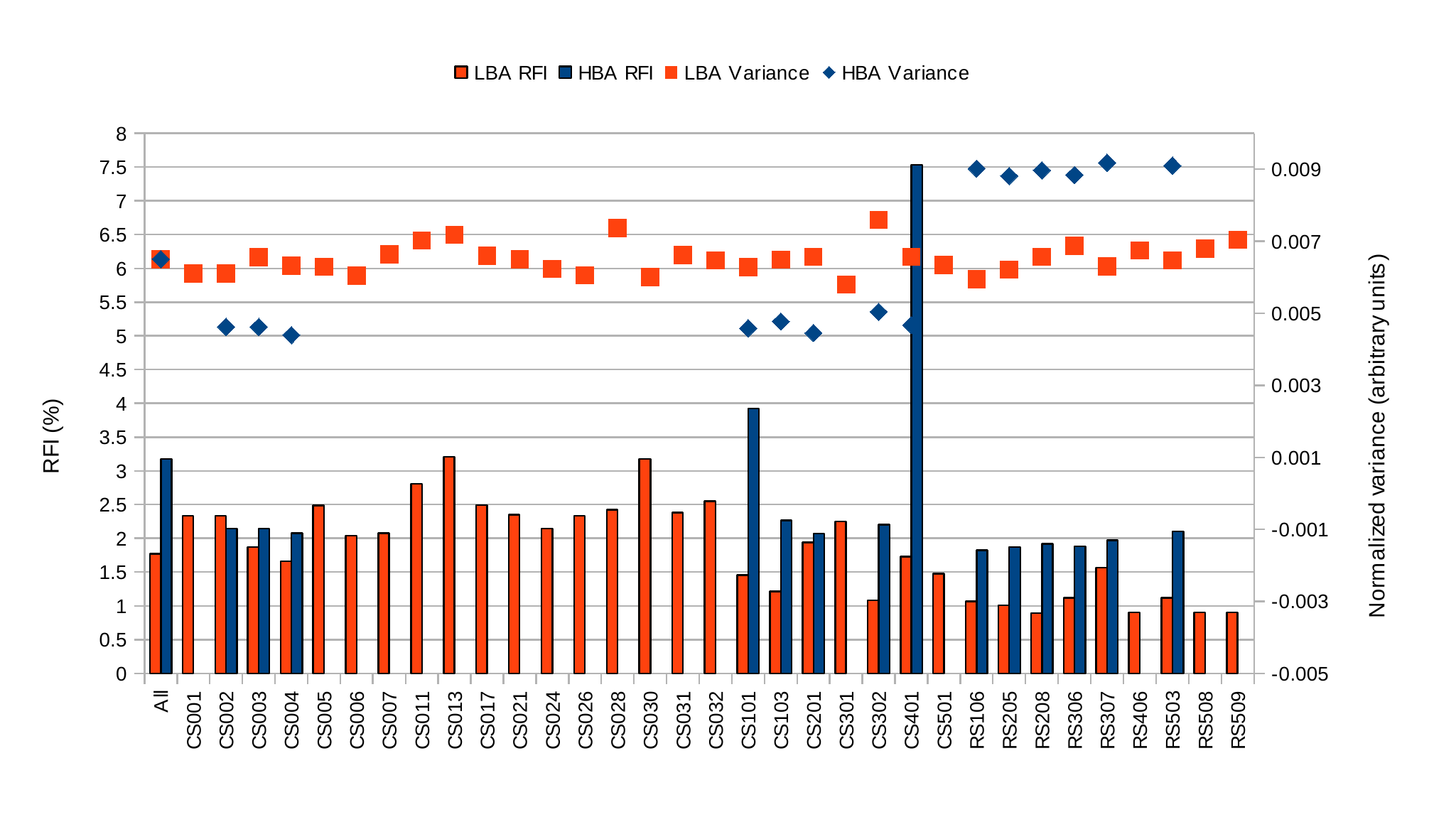}
\caption{The detected RFI percentages and the data variances per station, excluding auto-correlations.}
\label{fig:survey-antenna-rfi-plot}
\end{center}
\end{figure*}

\begin{figure}
\begin{center}
\noindent\includegraphics[width=24pc]{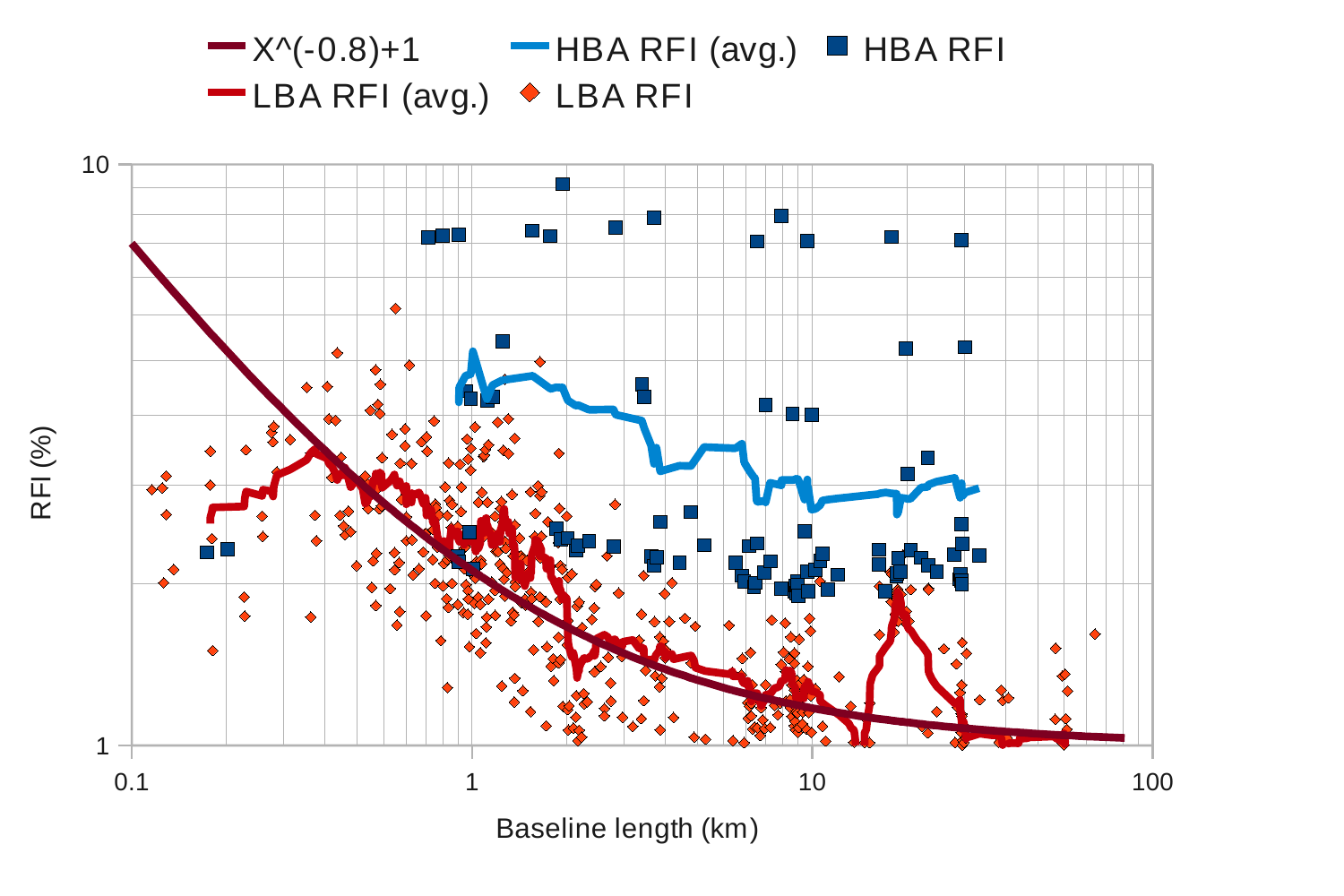}
\caption{RFI levels as a function of baseline length. Both axes are logarithmic. The dots represent the data (red: LBA, blue: HBA), while the lines show the trend of the points.}
\label{fig:survey-baselinelength-rfi-plot}
\end{center}
\end{figure}

\begin{figure*}[!htbp]
\begin{center}
\includegraphics[width=160mm]{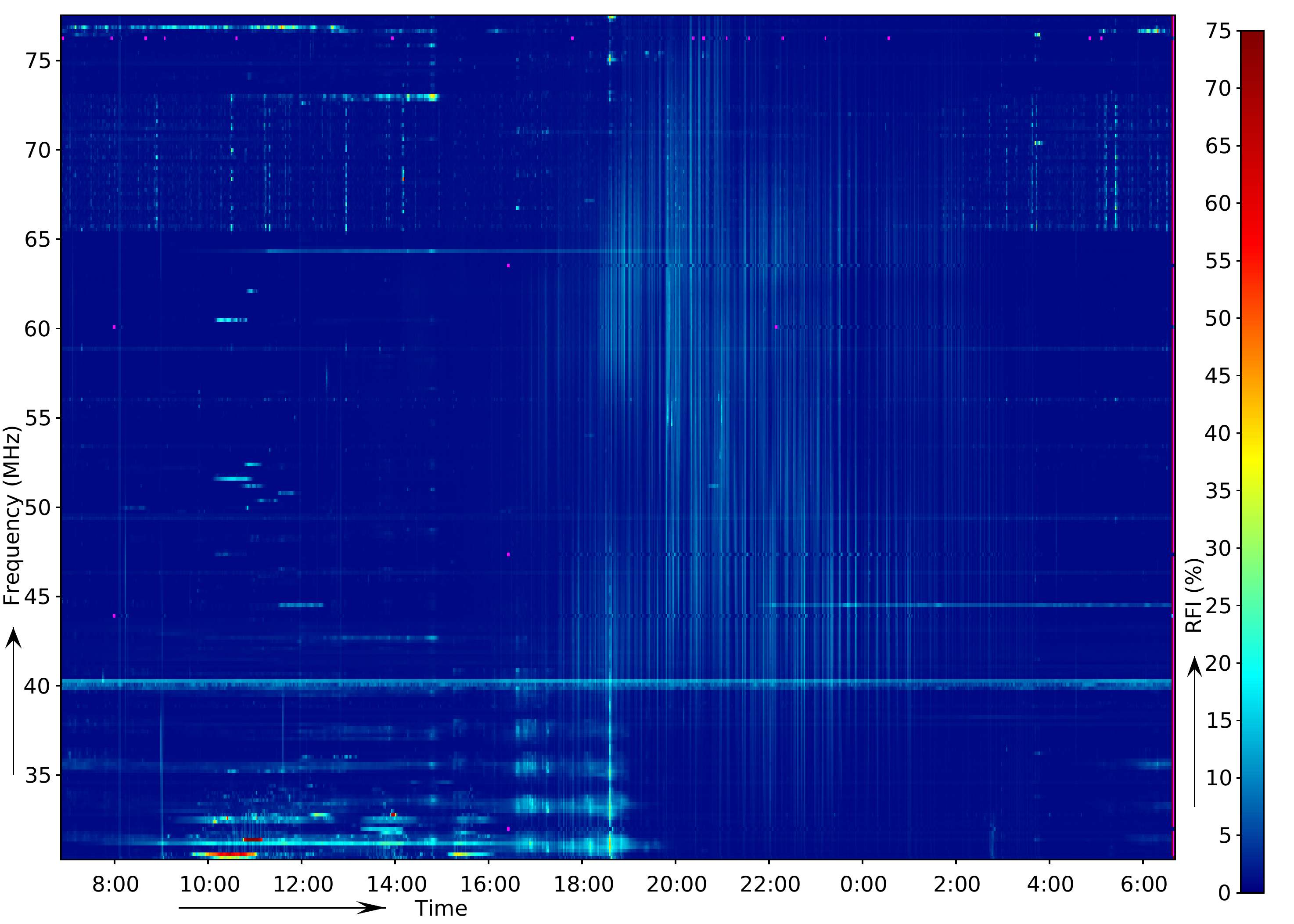}\vspace{2mm}
\includegraphics[width=160mm]{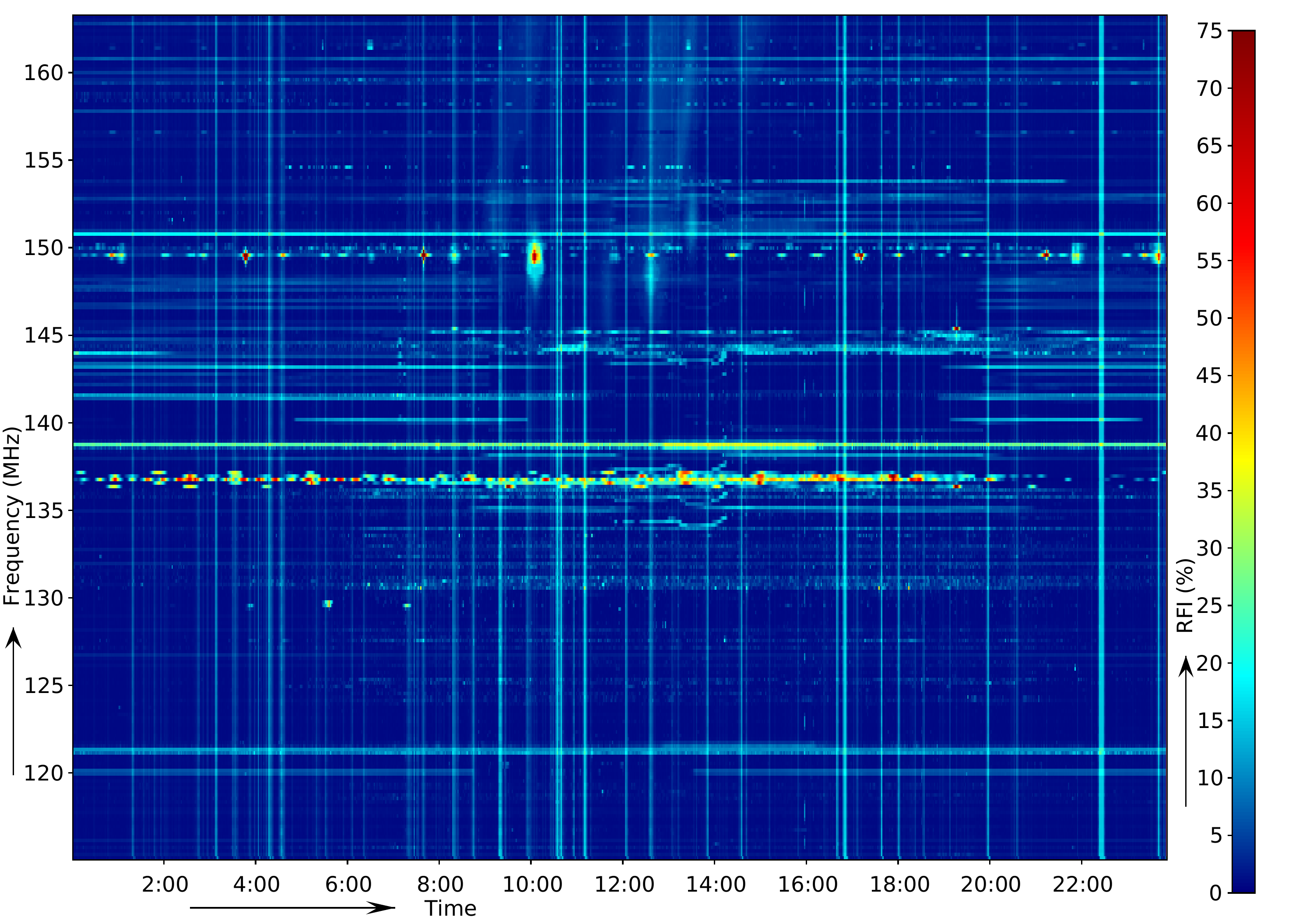}
\caption{Dynamic RFI occupancy spectrum for the surveys. Colour intensity represents the fraction of samples that were occupied in a specific time-frequency bin. The average over all baselines is shown. Top: LBA, bottom: HBA. The broad-band features in the LBA are likely to be ionospheric effects on Cas~A.}
\label{fig:survey-dynrfispectrum}
\end{center}
\end{figure*}

The default flagging pipeline found a total RFI occupancy of 2.24\% in the LBA survey at a resolution of 0.76~kHz and 1~s. However, we found that the flagger had a small bias. Because the sky temperature changes due to Earth rotation, the standard deviation of the data changes over time. The flagger applies a fixed sensitivity per sub-band and per baseline, and therefore does not take into account such changes over time. This is not an issue for short observations of about less than two hours during which the sky temperature does not change significantly. However, on long observations in which the sky temperature dominates the noise level, the flagger produces more false positives when sky temperature is higher and more false negatives when the sky temperature is lower.

Unfortunately, correcting for this effect requires an accurate estimate of the sky temperature, which in turn requires the interference to be flagged. Therefore, after the first flagging run, we have applied a second run of the flagger on normalised data. In the normalised data, each timestep was divided by the standard deviation of the median timestep in a window of 15 minutes of data, thereby assuming that the first run has removed the RFI. The calculation of the standard deviation per timestep was performed on the data from all cross-correlations. Therefore, this procedure results in a very stable estimate, although the cross-correlations of longer baselines will be less affected by the Galaxy, and this method will therefore not perfectly stabilise the variance in all baselines. In this article, when we refer to a ``second pass'' over the data, we refer to the above described second run of the flagger. Alternatively, it is also possible to calculate the standard deviation or median of differences over a sliding window during the first run and base the detection thresholds on this quantity, but this does not match well with the {\tt Sum\-Thresh\-old} method. The performance of the {\tt Sum\-Thresh\-old} method would significantly decrease when it can not process the data in one consecutive run with constant sensitivity. The {\tt Sum\-Thresh\-old} method is crucial for the accuracy of the flagger.

After having corrected for the changing sky temperature, the detected RFI occupancy is 1.77\%. The RFI occupancy over frequency is plotted in Fig.~\ref{fig:survey-rfi-freq-plot}, while Fig.~\ref{fig:survey-antenna-rfi-plot} shows the percentages of flagged data per station. The stations with higher station numbers are generally farther away from the core, and therefore provide longer baselines. The remote stations (RS) are farthest away and for these stations, the high-band antennas are not split into two sub-stations. Fig.~\ref{fig:survey-antenna-rfi-plot} shows that the stations closer to the core generally have a higher RFI occupancy. This can be explained by the larger number of short baselines in the central fields and the fact that RFI is decorrelated on the longer baselines. By plotting the RFI as a function of baseline length as shown in Fig.~\ref{fig:survey-baselinelength-rfi-plot}, it is observed that the RFI decreases as a function of baseline length for lengths $>$ 300 m, and closely follows a power law that asymptotically reaches $\sim$1.0\%. This asymptote might be reached because of false positives and interfering sources such as satellites that do not decorrelate in the longer baselines.

Statistics in this paper are all based on cross-correlations.
Detailed RFI statistics for the auto-correlations are not presented.
Nevertheless, visual inspection of the auto-correlations show
stronger RFI contamination and higher RFI incidence compared to the
cross-correlations. Auto-correlations are typically not used for imaging
or in EoR
angular power spectrum measurements. However, a total power experiment
using auto-correlations to detect signals from the EoR is underway, and
results from pilot observations, including RFI statistics, are in
preparation (Vedantham et al., priv. com., 2012).

\begin{figure*}
\begin{center}
\hspace*{-0.3cm}\includegraphics[width=35pc]{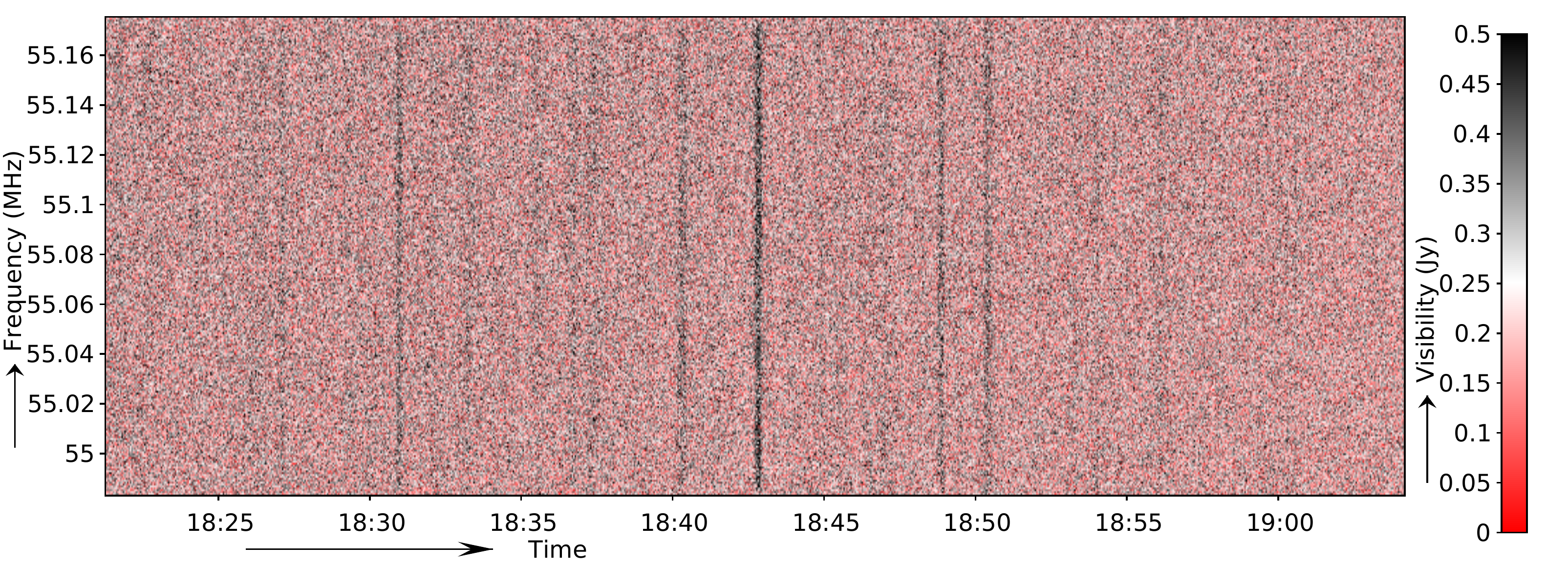}
\caption{Data from the LBA 4-km long baseline CS001 $\times$ RS503 at high frequency resolution, showing strong fluctuations of 1--10~s. The flagger detects these as RFI.}
\label{fig:survey-lba-spikes}
\end{center}
\end{figure*}

The LBA set contains many broadband spikes between 18:00 and 0:00~UTC. These are detected by the flagger as RFI, and are therefore visible in the dynamic RFI occupancy spectrum of Fig.~\ref{fig:survey-dynrfispectrum}. An example of the spikes at high resolution on a 4~km baseline is shown in Fig.~\ref{fig:survey-lba-spikes}. Individual spikes affect all samples for 1--10 seconds. Despite the relatively long baseline of 4~km, these spikes have evidently not yet become incoherent. On the 56 km baseline CS001~$\times$~RS509, the spikes are not visually present in the time-frequency plot, but some of them are still detected by the flagger because of an increase in signal to noise in these timesteps. It is assumed that they are strong ionospheric scintillations of signals from Cassiopeia~A, because they correlate with its apparent position. Cas.~A is 32$\degree$ away from the NCP, which is the phase centre. Cygnus~A might also cause such artefacts, but is 50$\degree$ from the phase centre.

\begin{figure*}
\begin{center}
\includegraphics[width=130mm]{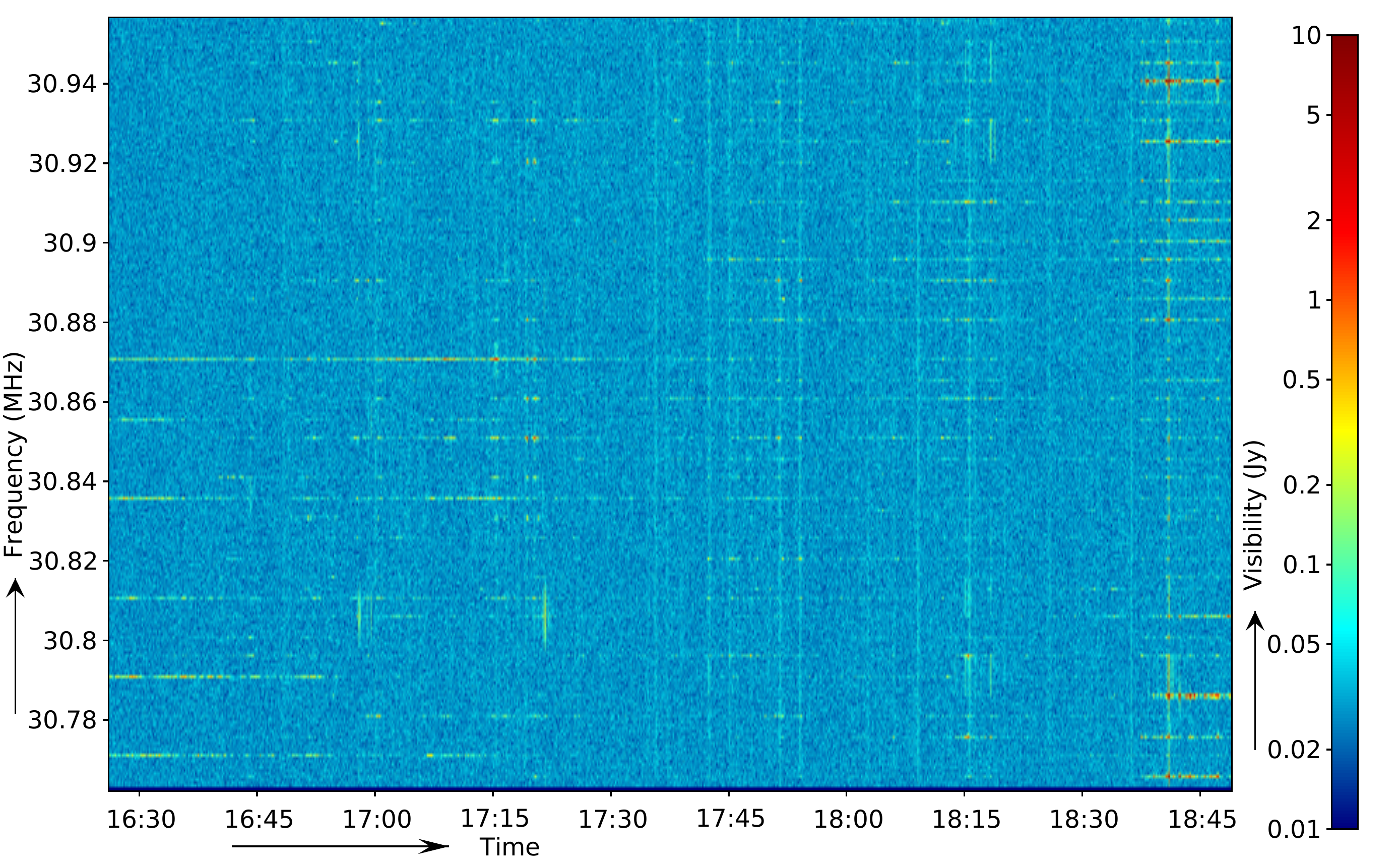}
\includegraphics[width=130mm]{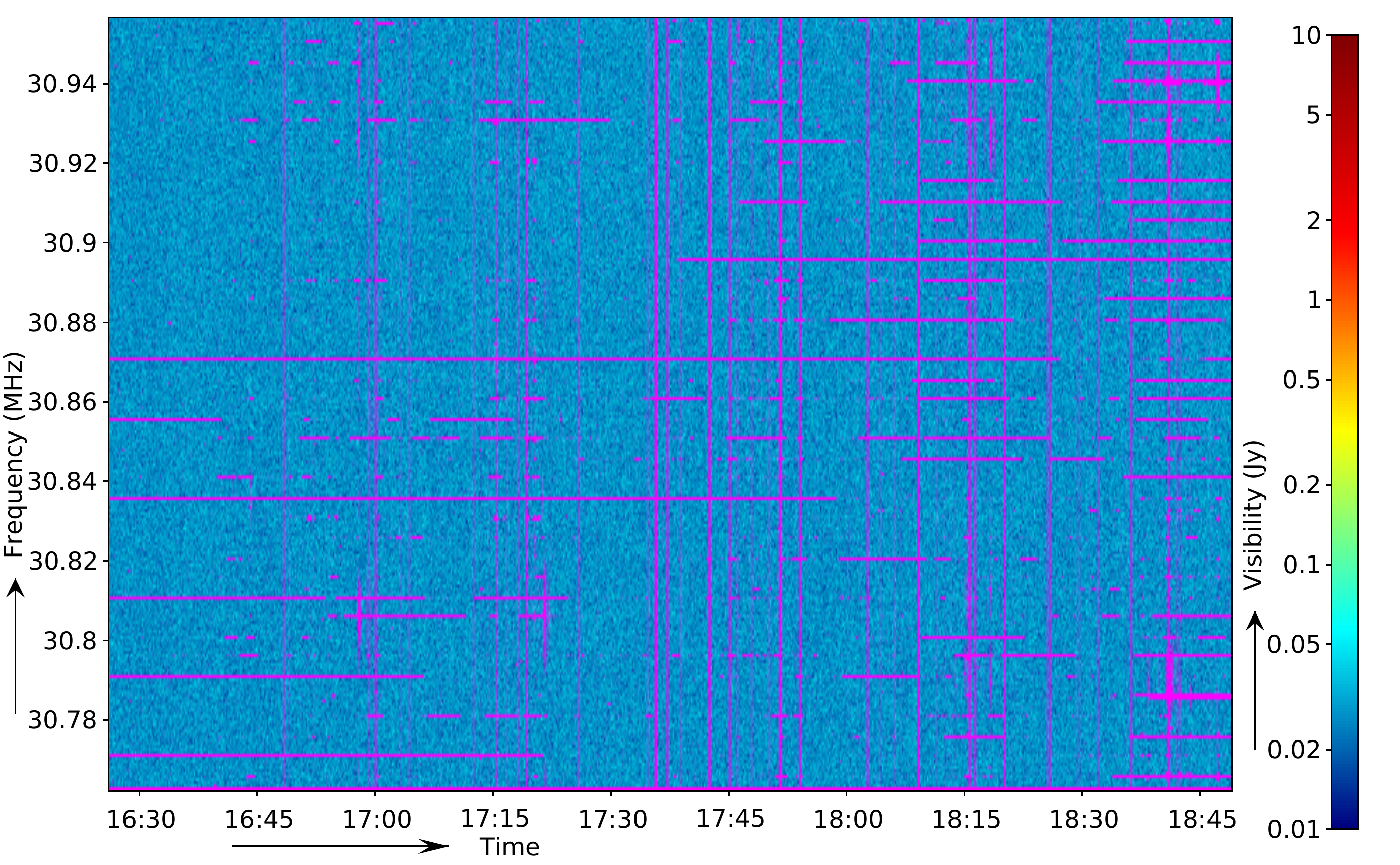}
\caption{A dynamic spectrum of data from one sub-band of the LBA survey, formed by the correlation coefficients of baseline CS001 $\times$ CS002 at the original frequency resolution of 0.76 kHz. The displayed sub-band is one of the most affected sub-bands in terms of the detected level of RFI. The top image shows the original spectrum, while the bottom image shows with purple what has been detected as interference. }
\label{fig:survey-sb004-high-rfi-level-example}
\end{center}
\end{figure*}

At the very low frequencies, around 30~MHz and 17:00--18:00~UTC, a source is visible that shows many harmonics. A high resolution dynamic spectrum is shown in Fig.~\ref{fig:survey-sb004-high-rfi-level-example}. It is likely that this source has saturated the ADC or amplifiers. Nevertheless, its harmonics are flagged accurately, and it causes no visible effects in the cleaned data.

\subsection{HBA survey}
The analysis of the HBA survey shows a higher RFI occupancy of 3.18\%. The increased artefacts in the RFI occupancy spectrum of the HBA in Figs.~\ref{fig:survey-rfi-freq-plot} and \ref{fig:survey-dynrfispectrum} also confirm that the HBA is more contaminated by interference than the LBA. However, as can be seen in Fig.~\ref{fig:survey-antenna-rfi-plot}, almost all stations have less than 2.5\% RFI occupancy. Stations CS101HBA0 and CS401HBA0 are the only two exceptions, with respectively 3.9\% and 7.5\% RFI, and are also a cause of the higher level of RFI compared to the LBA survey. Despite the larger fraction of RFI in stations CS101HBA0 and CS401HBA0, the data variances of these are similar to the other stations. This suggests therefore the presence of local RFI sources such as a sparking electric fence or a lawn mower near these two stations, which have successfully been excised by the flagger. This RFI source seems to have been temporary, as recent observations show normal RFI detection occupancies of less than 3\% for data from this station. Fig.~\ref{fig:survey-antenna-rfi-plot} also shows that the variances of the remote stations are higher. This is because these stations contain twice as many antennas.

As in the case of the LBA survey, detected RFI occupancies in the HBA are affected by the changing sky temperature. Again we have performed a second pass in which the normalised data was flagged. However, because the HBA system is far less sky noise dominated than the LBA system \citep{lofar-antenna-performance}, the noise level in the HBA data is less affected by the changing sky. Consequently, the difference between the first and second pass is minor, and after the second pass the detected level of RFI is less by only 0.04\%.

In Fig.~\ref{fig:survey-baselinelength-rfi-plot}, for the HBA it is harder to assess whether the level of RFI decreases significantly on longer baselines due to the smaller number of baselines.

\begin{figure*}
\begin{center}
\includegraphics[height=76mm]{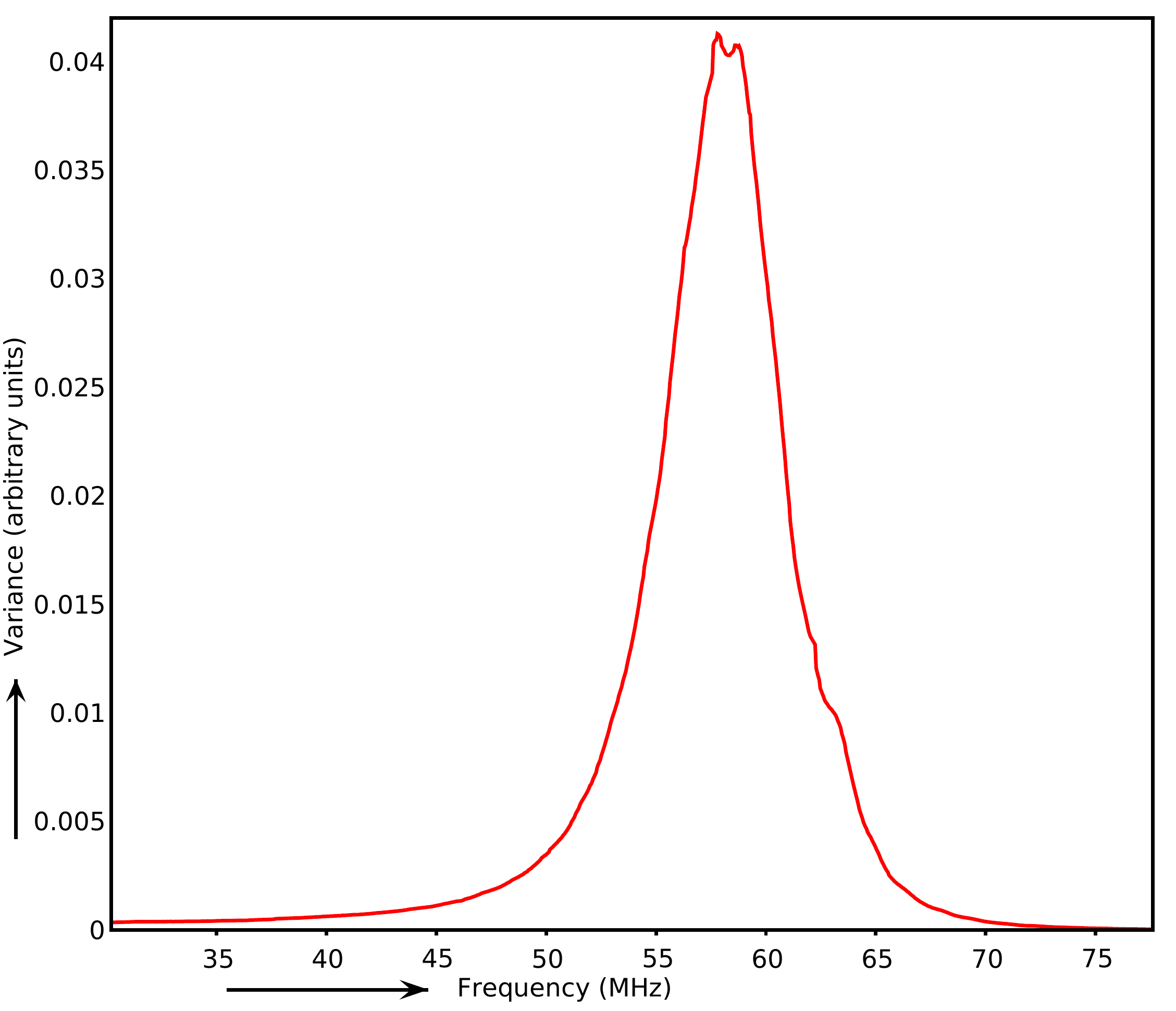} \hspace{2mm}
\includegraphics[height=76mm]{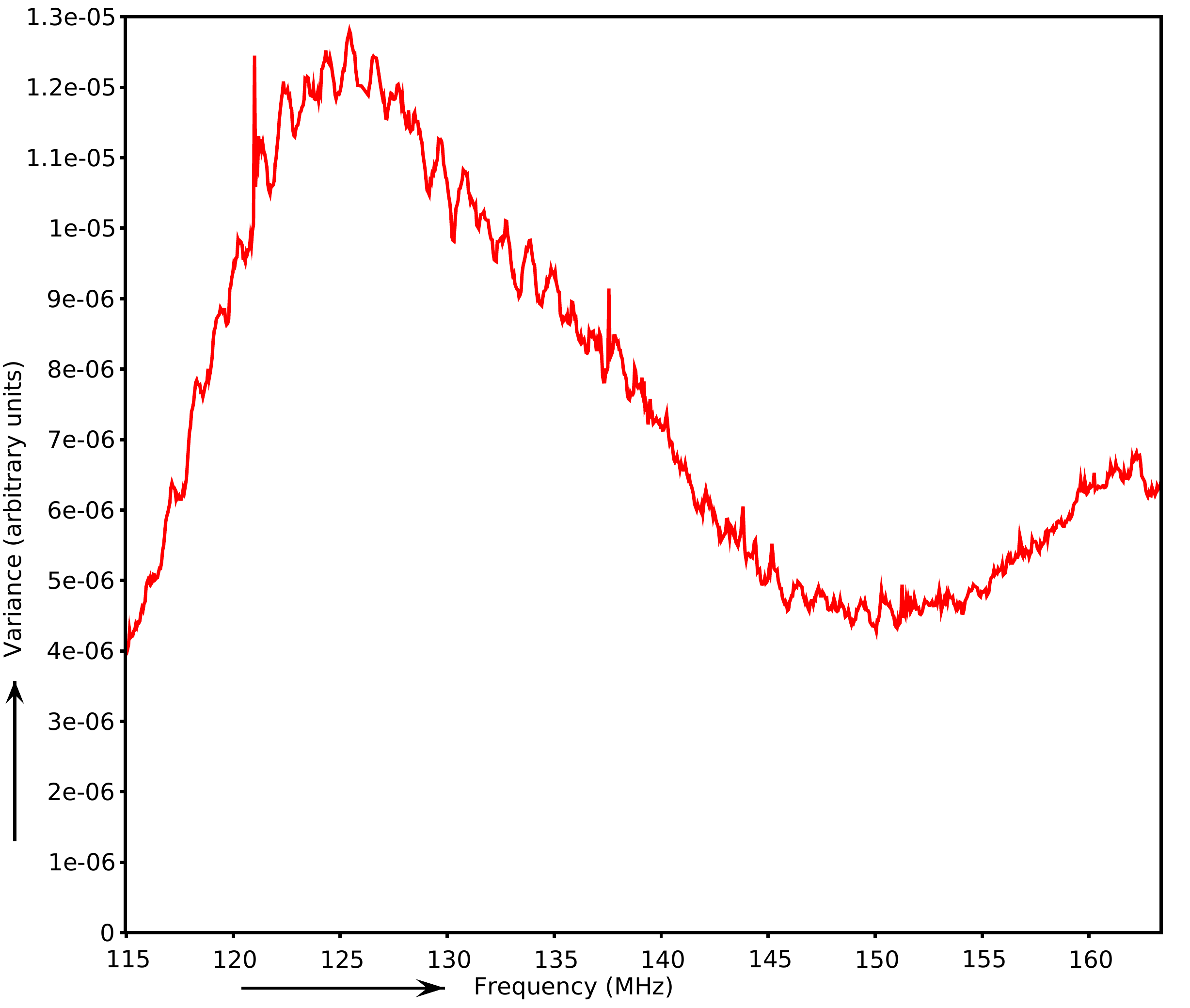}
\caption{The post-flagging spectra of data variances for both RFI surveys. The dominating effect is the antenna frequency response. In the HBA (right plot), a strong ripple of around 1~MHz is apparent, which is caused by reflections in the antenna cables.}
\label{fig:survey-variance-freq-plot}
\end{center}
\end{figure*}

\begin{figure}
\begin{center}
\hspace*{-5mm}\includegraphics[width=98mm]{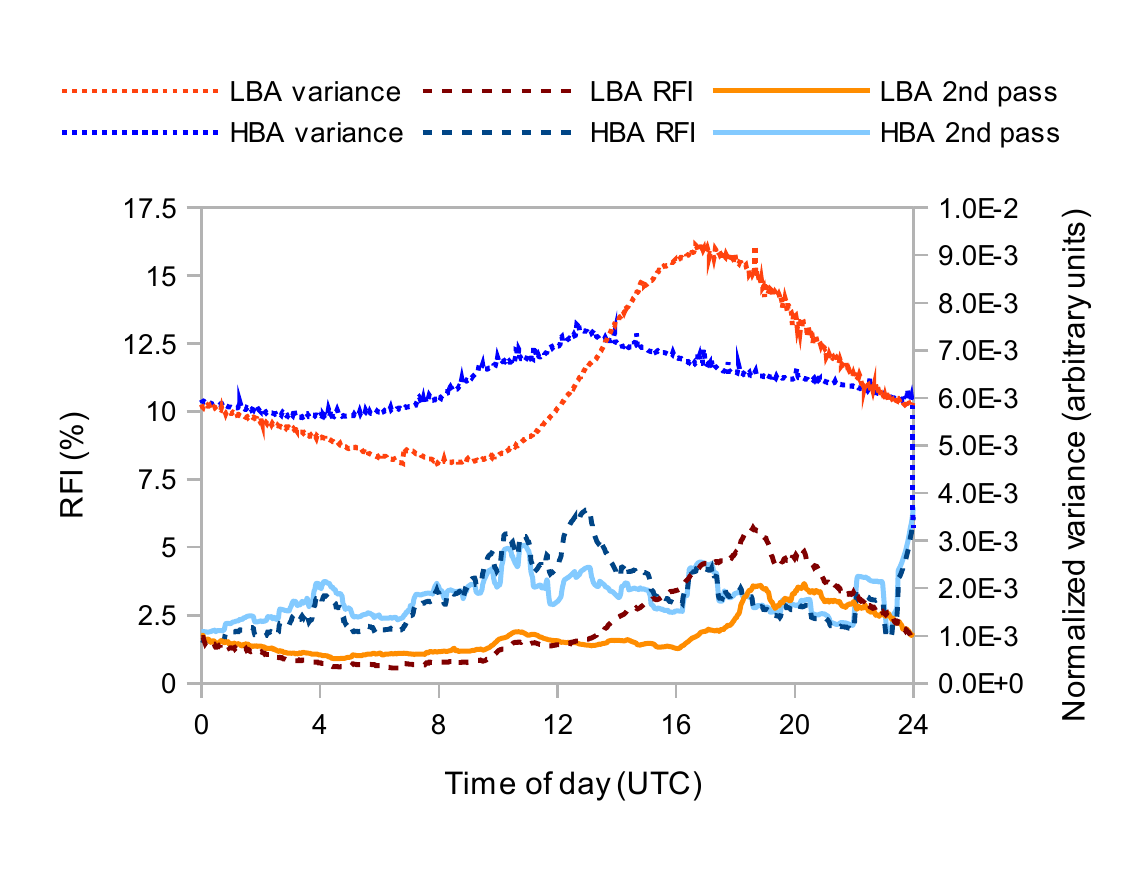}
\caption{RFI levels and variances as a function of the time of day. The RFI percentages are smoothed. Although there is some variation in the detected RFI during the observation, this is likely not because of a different occupation of RFI between day and night. Instead, they are likely caused by the changing sky, since they correlate with the variance of the data and visual celestial artefacts in the dynamic spectra.}
\label{fig:survey-hour-of-the-day-plot}
\end{center}
\end{figure}

\begin{figure*}
\begin{center}
\noindent\includegraphics[height=110mm]{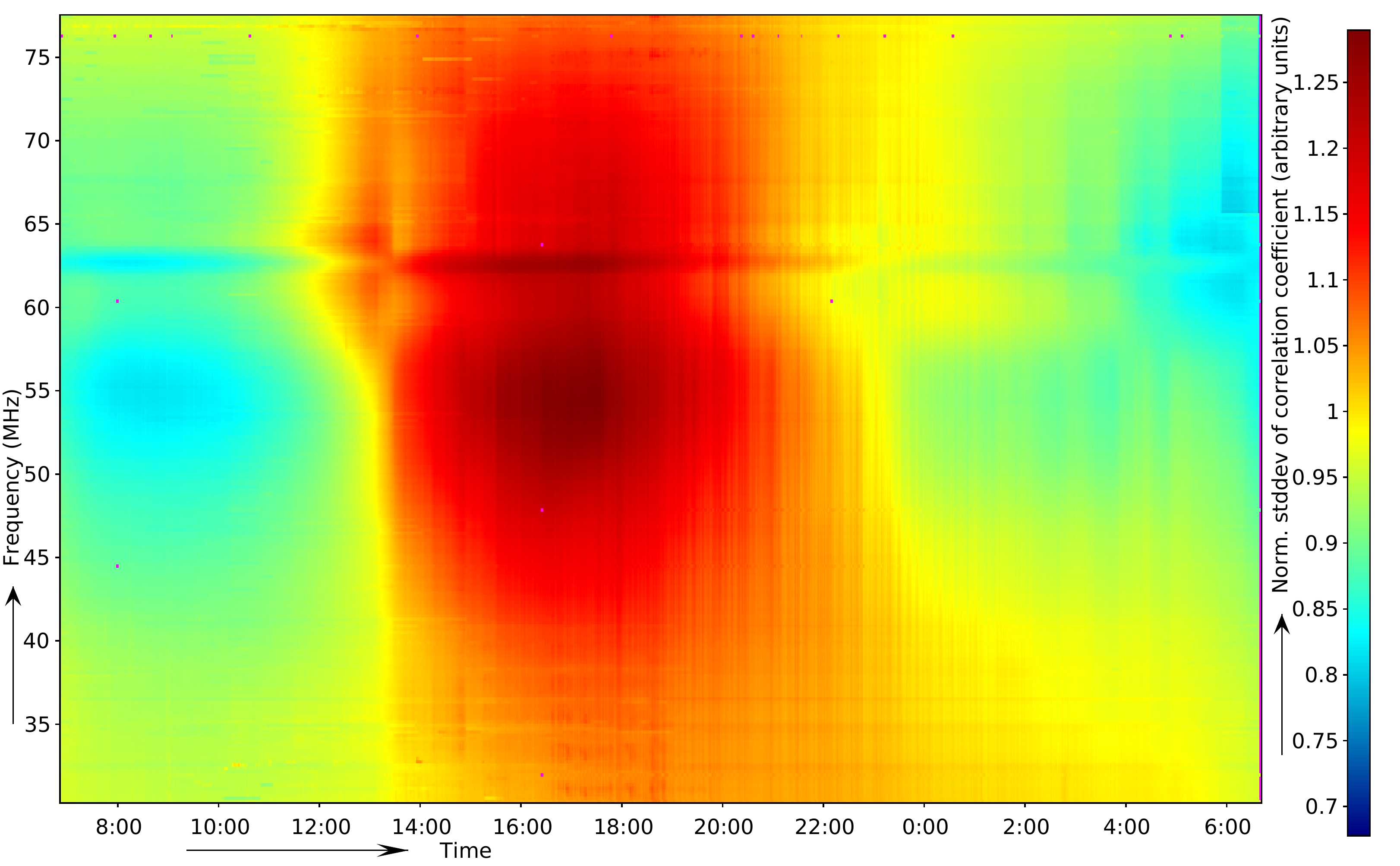}
\noindent\includegraphics[height=110mm]{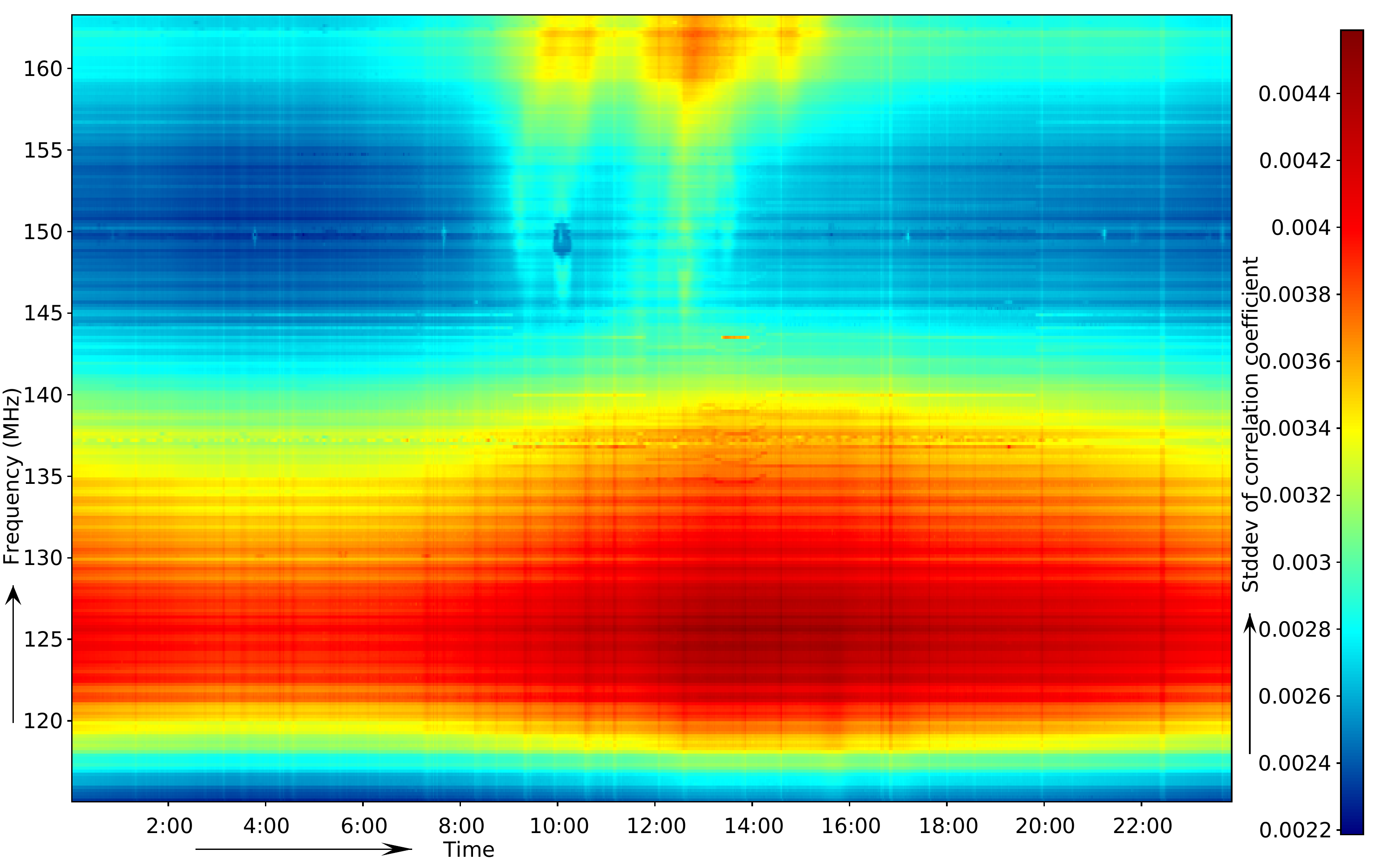}
\caption{The standard deviation over time and frequency during the surveys. In the LBA set, the individual statistics of each sub-band were divided by the Winsorized mean of the sub-band, to correct for the antenna response on first order. In the LBA set, no residual RFI is visible, except some weak residuals near the edges of the band. A few purple dots can be seen in the data, which denotes missing data. The HBA set shows a bit more undetected, but weak RFI residuals.}
\label{fig:survey-stddev-var-spectrum}
\end{center}
\end{figure*}

\subsection{Overall results}
After the automated RFI detection, there are generally no harmful interference artefacts in the data at the level at which we make images at the moment. The variance over frequency and time are displayed in respectively Fig.~\ref{fig:survey-variance-freq-plot} and Fig.~\ref{fig:survey-hour-of-the-day-plot}, and are displayed in a time-frequency diagram in Fig.~\ref{fig:survey-stddev-var-spectrum}. While the HBA variances look clean in most frequencies, there are a few spikes of RFI that evidently have not been detected. These look like sharp features in the full spectrum, but are in fact smooth features when looking at full resolution. Because they are smooth at the raw sub-band resolution, the flagger does not detect them as RFI. Although there are interference artefacts visible in the HBA spectrum, after detection the data can be successfully calibrated and imaged. A possible second stage flagger to remove any residual artefacts will be discussed in \S\ref{sec:env-discussion}. The LBA variances show only a few RFI artefacts around its higher frequencies.

The HBA spectrum contains a clearly visible ripple of about 1 MHz. This has been identified as the result of reflection over the cables, resulting from an impedance mismatch in the receiver unit. In fact, a similar phenomenon occurs in LBA observations, but because of the steeper frequency response and because not all LBA cables are of the same length, it is less apparent. The reflection is also less strong in the LBA, due to the better receiver design. A Fourier transform of the LBA variance over frequency shows slight peaks at twice the delays of the cables.

\subsection{Day and night differences}
One might expect a lower RFI occupancy during the night, i.e., during 23:00--6:00~UTC (Local time is UTC+1). We use Fig.~\ref{fig:survey-hour-of-the-day-plot} to assess this possibility. The figure shows variance and RFI occupancy as a function of the hour of the day in UTC. However, after one pass of flagging, the data are highly dominated by the changing sky. Moreover, the LBA data also contain artefacts due to Cassiopeia~A, which causes some spikes in the data due to strong ionospheric scintillation between 18:00 and 0:00~UTC.

Unfortunately, the biasing effect of the sky temperature is not completely removed even with a second pass over the data. There is no significant additional trend visible. This implies that there is no significant relation between the hour of the day and the RFI occupancy due to less activity at night. This is also evident in the dynamic spectra of RFI in Fig.~\ref{fig:survey-dynrfispectrum}, which show no obvious increase or decrease of transmitters during some part of the day, and many transmitters start and end at random times. In a few cases, the starting of a transmitter at a certain frequency coincides with the termination of a transmitter at a different frequency, suggesting that some transmitters hop to another frequency. In Fig.~\ref{fig:survey-dynrfispectrum}, such transmissions can be seen between 140 and 145~MHz. These transmissions end at 9:00~UTC, while at the same time several transmissions start around 135--140~MHz.

To further explore the possibility of increased RFI during daytime of the HBA set, we have performed the same analysis on a 123--137~MHz subset of the HBA observation. There are two reasons that the difference between day and night might be better visible in this frequency bandwidth: (i) all the visual peaks of detected RFI that correspond to the Sun have a frequency higher than 145 MHz; and (ii) this band corresponds to air traffic communication, which is less used during the night. Nevertheless, we still do not see a significant increase of RFI in this subset of the data.

In summary, any effect of increased activity during the day is not significant enough to be identifiable in the detected occupancies of either the LBA or the HBA data set. The post-flagging data variances are dominated by celestial effects, i.e., the Sun, the Milky Way or Cassiopeia~A, and contain no clear signs of a relation between day and night time either.

\subsection{Resolution \& flagging accuracy}
\begin{figure}
\begin{center}
\noindent\includegraphics[width=90mm]{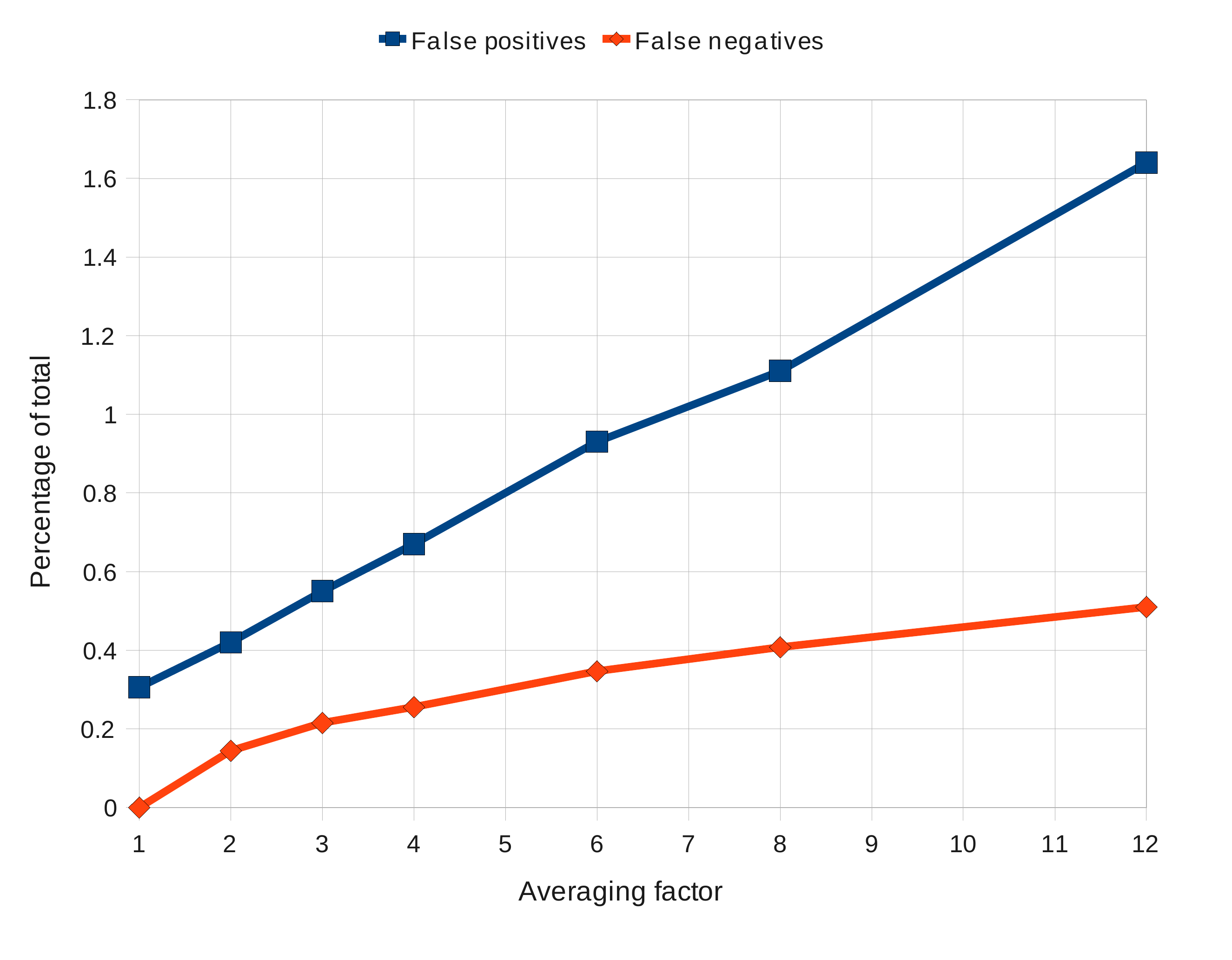}
\caption{This plot shows the RFI detection accuracy as a function of frequency resolution, using data from the LBA survey. The frequency resolution is 0.76 kHz at an averaging factor of 1. The resolution is lowered by averaging the samples in adjacent channels. The time resolution is fixed at 1 s.}
\label{fig:frequency-averaging}
\end{center}
\end{figure}
The frequency and time resolution of observations do affect the accuracy of the interference detection. It is, however, not known how significant this effect is. To quantify this, we have decreased the frequency resolution of the HBA RFI survey in several steps and reflagged the averaged set. Subsequently, the resulting flags were compared with the flags that were found at high resolution. The original high resolution flags were used as ground truth. 

We found that the level of false positives is approximately linearly correlated with the decrease in resolution. Unfortunately, false positives cause samples in our ground truth to be misclassified as RFI, and will therefore show up as false negatives in the lower resolution detections. Therefore, the false positives for the ground truth data were determined by extrapolating the false-positives curve of the sets with decreased resolution. This yields a false-positives rate of 0.3\%, which subsequently has been subtracted from the false negatives. The resulting curves after these corrections are plotted in Fig.~\ref{fig:frequency-averaging}.

Because the test is computationally expensive, we have not performed the same test on the LBA survey or for the time resolution. However, tests on small parts of the data show that decreasing the time resolution results in similar false-negatives curves compared with decreasing the frequency resolution, although it causes about 20\% less false positives. Therefore, from the RFI detection perspective, it is slightly better to have higher frequency resolution compared to higher time resolution at LOFAR resolutions. It is still to be ascertained whether the small amount of data was representative enough to draw generic conclusions.

\subsection{False-positives rate}
If we assume that the least contaminated sub-bands in Fig.~\ref{fig:survey-dynrfispectrum} are completely free of RFI on the long baselines, they can be used to determine the false-positive rate of the flagger. For the LBA set, we selected the 4-km long baseline CS001~$\times$~RS503 and the 56-km long baseline CS001~$\times$~RS509 of one the best centre sub-bands at 55~MHz. For the 4-km baseline the total detected fraction of RFI is 0.75\%, while for the 56-km baseline it is 0.73\%. However, the 4-km baseline contains some broadband spikes around 18:40~h, as shown in Fig.~\ref{fig:survey-lba-spikes}. On the 56-km baseline CS001~$\times$~RS509, the spikes can not be seen in the time-frequency plot, but some of them are still detected by the flagger because of an increase in signal to noise in these timesteps.

To get a more accurate estimate of the base level of false positives, we have also determined the false-positives rate by using only the last 50~min of the sub-bands. Visual inspection of this data shows indeed no RFI, except for two timesteps in the 4~km baseline that might have been affected, but these can not be assessed with certainty. The flagger does flag those timesteps, hence we ignore them in the analysis. When flagging only the 50 minutes of 4 km baseline data, thereby making sure that the threshold is based only on this 50 min of data, a fraction of 0.6\% was flagged. If one assumes that the selected data contains no other RFI, then this value is the rate of falsely flagged samples. In the 56 km baseline, the same analysis leads to a slightly lower rate of false-positives of 0.5\%.

The 0.6 and 0.5\% detection rates are the result of flagging on all four cross-correlations (XX, XY, YX and YY). In the samples that have been detected as RFI, we observe that there are zero samples flagged in more than one cross-correlation for that particular time and frequency, thus they are completely uncorrelated. Each cross-correlation adds independently about 0.13--0.15\% of falsely detected samples. In a simulated baseline with complex Gaussian noise the flagger detects 0.14\% as RFI, thus these values are similar to the expected ones.

Estimating the false-negatives rate is harder to carry out, because we do not know the exact interference distribution. Because there are almost no RFI artefacts after flagging, the false-negatives can be assumed to be insignificant in most cases.

\section{Comparison with other observations} \label{sec:env-compare}
Although we have analysed a substantial amount of survey time, it is useful to validate whether the two observations are representative samples for determining the LOFAR interference environment. Unfortunately, comparing the surveys with other observations is hard at this point, because often during LOFAR commissioning observations are being carried out with lower frequency and time resolutions to reduce the data size, and the analysed 24~h surveys are the only substantial observations performed at the standard LOFAR resolution. A relative comparison can still be done for lower resolution data. There are no strong sources in the targeted NCP field, which further complicates the comparison. Fields that do have strong sources might trigger the flagger more easily, yielding higher detection rates.

\begin{table*}\begin{center}
\begin{threeparttable}
\caption{Observations and their RFI occupancy as reported by automated detection. The bold entries are the surveys analysed in this article.}\label{tbl:eor-obs-statistics}
%\begin{center}
\begin{tabular}{cccccccr}
\hline
\hline
\textbf{Date} & \textbf{Start (UTC)} & \textbf{Duration} & \textbf{Id} & \textbf{Target} & $\Delta\nu$ (kHz) & $\Delta t$ (s) & \textbf{RFI}\tnote{[1]} \\
\hline
\multicolumn{8}{c}{\emph{LBA observations (frequency range $\approx$ 30 -- 78 MHz)}}\\
\hline
2010-11-20 & 19.33 & 5 min & L21478 & Moon & 3.0 & 1 & 4.6\% \\
2010-11-20 & 19.43 & 6 h   & L21479 & Moon & 3.0 & 1 & 10.3\% \\
2011-04-14 & 19.00 & 8 h   & L25455 & Moon & 0.76 & 1 & 4.3\% \\
%2011-05-04 & 23.00 & 18 h  & L25937 & Cas~A & 3.0 & 1.1\% \\ <-- something went wrong during this observation
\textbf{2011-10-09} &  \textbf{6.50} & \textbf{24 h}  & \textbf{L31614} & \textbf{NCP} & \textbf{0.76} & 1 & \textbf{1.8}\% \\
\hline
\multicolumn{8}{c}{\emph{HBA observations (frequency range $\approx$ 115 -- 163 MHz)}}\\
\hline
2010-11-21 & 20.26 & 5 min & L21480 & Moon & 3.0 & 1 & 5.6\% \\
\textbf{2010-12-27} &  \textbf{0.00} & \textbf{24 h}  & \textbf{L22174} & \textbf{NCP} & \textbf{0.76} & 1 & \textbf{3.2}\% \\
2011-03-27 & 20.00 & 6 h   & L24560 & NCP  & 3.0 & 2 & 1.5\% \\
2011-04-01 & 16.08 & 6 h   & L24837 & 3C196 & 3.0 & 2 & 2.6\% \\
2011-06-11 & 11.30 & 1.30 h & L28322 & 3C196 & 3.0 & 2 & 6.5\% \\
2011-11-17 & 18.00 & 12 h & L35008 & NCP & 3.0 & 2 & 3.6\% \\
2011-12-06 & 2.36 & 25 min & L36691 & 3C196 & 3.0 & 2 & 5.5\% \\
2011-12-06 & 8.34 & 25 min & L36692 & 3C295 & 3.0 & 2 & 8.0\% \\
2011-12-20 & 7.39 & 30 min & L39562 & 3C295 & 3.0 & 2 & 2.5\% \\
2012-01-26 & 2.00 & 5.30 h & L43786 & 3C295 & 3.0 & 2 & 3.6\% \\
% &  &  & L & & 3.0 & \% \\
\hline
\hline
\end{tabular}
%\end{center}
Notes:
\begin{tablenotes}\begin{small}
\item [{[1]}] RFI occupancy as found by automated detection. For some targets, this is too high because of the band-edge issues that are discussed in the text, leading to approximately a 1--2\% increase in 3-kHz channel observations.
\end{small}
\end{tablenotes}
\end{threeparttable}\end{center}
\end{table*}

To assess the differences between different observations, we have performed detection occupancy analysis of several other observations. For this purpose, we collected several LOFAR observations that were used for quality assessment. These were subsequently processed similarly to how we processed the surveys. The observations were selected independent of their quality, hence they sample the RFI situation randomly. However, it is important to note that in our experience the data quality, such as the achieved noise level of the final image, is quite independent of the detected RFI occupancy. Much more relevant is the position of the Sun in the sky, the state of the ionosphere and the stability of the station beam. These have very little effect on the detected RFI occupancy.

Table~\ref{tbl:eor-obs-statistics} lists these other observations and shows their statistics. The number of involved stations varies between the observations, but as many as possible core stations were used in all observations.

Currently, there is an issue with some LOFAR observations that causes higher RFI detection rates in fields with strong sources. This is caused by the edges of sub-bands in some cross-correlated baselines. These edges are flagged because they show time-variable changes that are very steep in the frequency direction. This effect is only observed in cross-correlations that involve exactly one Superterp station, so it is assumed that this is a bug in the station beamformer or correlator. In 64 channel observations that show this issue, the first and last sub-band channels get flagged in about half of the baselines, leading to about a 1--2\% higher detected RFI occupancy. The issue only arises in fields that contain strong sources, and is consequently not affecting the 24~h RFI surveys, because there are no such sources in the NCP field. All 3C196, 3C295 and Moon observations do show the issue.

The average detected RFI occupancies are 5.4 and 4.3\% with standard deviations 3.5 and 2.0\% for the LBA and HBA observations respectively. Therefore, it appears that the analysed 24~h RFI surveys, with 2.4 and 3.2\% RFI occupancy in the low and high bands respectively, are less affected by RFI than the average observation. If one however assumes that the observations with lower time and frequency resolutions have an approximately 1.0\% RFI increase, which seems to be a reasonable estimate according to Fig.~\ref{fig:frequency-averaging}, and taking into account that the subband-edge issue causes another 1.5\% RFI increase on average in the fields with strong sources, the averages after correction for these effects become 3.7 and 2.4\%. Therefore, the RFI occupancies of the 24~h surveys seem to be reasonably representative for the RFI occupancy of LOFAR at its nominal resolution of 0.76~kHz with 1~s integration time. On the other hand, it also shows that $3$~kHz channels may well suffice for regular LOFAR observations.

Visual inspection of the same data agreed with this observation: the RFI environment is not significantly different between different observations. The only exception was the Moon observation of 2010-11-20, which seems to contain unusual broadband interference over the entire duration of the observation. Note that the moon is known to reflect some of the RFI, but such reflections are too faint to trigger the flagger. The shape and frequency at which the interference occurred is not like in any other observation. Therefore, we suspect that either something went wrong during this particular observation or ionospheric conditions were exceptional. According to weather reports, it was observed at the day with highest humidity of the year, although we have no explanation why this would influence the RFI occupancy.

\section{Discussion \& conclusions} \label{sec:env-discussion}
We have analysed 24-h RFI surveys for both the high-band and low-band frequency range of LOFAR. Both sets show a very low contamination of detectable interference of 1.8 and 3.2\% for the LBA and HBA respectively. In the considered frequency ranges, these are predicted to be representative quantities for what can be expected when LOFAR starts its regular observing with resolutions of 0.76~kHz and 1~s. Therefore, the LOFAR radio environment is relatively benign, and is not expected to be the limiting factor for deep field observing. However, it remains important that the spectrum is not used for broadband transmitters such as DAB stations. Also strong local interference can become a problem. For example, it is currently not clear what the effect of windmills close to the LOFAR stations might be, since these can potentially reflect and generate additional and time-varying interference. We have also not considered LOFAR's entire frequency range, but instead focused on the most sensitive region. This region is probably the least contaminated by RFI, because the RFI situation is worse below 30~MHz and above 200~MHz. We have focused on the RFI situation for imaging observations. The RFI situation might be different when observing with a much higher time resolution, as is done for the LOFAR transient key science project.

Almost all visible interference is detected after the single flagging step at highest resolution, and RFI that leaks through is very weak. This agrees with the first imaging results, which are thought to be limited by beam and ionospheric calibration issues and system temperature, but not by interference. However, whether this will still be the case for long integration times of tens of nights, as will be done as part of the Epoch of Reionization project, remains to be seen. In that case, one might find that weak, stationary RFI sources add up coherently, and might at some point become the limiting factor. Nevertheless, the situation looks promising: our first-order flagging routines use only per-baseline information, but remove in most cases all RFI that is visible in the spectra. The resulting integrated statistics of 24 hours show very few artefacts of interference, and these are causing no obvious issues when calibrating and imaging the data.

If RFI does become a problem, there are many methods at hand to further excise it. The interference artefacts still present can be flagged with a second stage flagger. In such a stage, the flagger could use the information from the entire observation, and such a strategy would be more sensitivity for weak stationary sources. Moreover, the Fourier transform used for imaging is a natural filter of stationary interference. Without fringe stopping, a single baseline will observe a stationary source as a constant source. Therefore, the contribution of stationary sources would end up at the North Pole. With sufficient uv-coverage, the sidelobe of this source at the NCP will be benign. Furthermore, if necessary these can be further attenuated with filtering techniques, such as low-pass filters that remove contributions in the data with a fringe frequency faster than can be generated by on-axis sources \citep{post-correlation-filtering}. Therefore, we believe that RFI will not keep LOFAR from reaching its planned sensitivity.

Unexpectedly, we found that the RFI occupancy is not significantly different between day and night. In both the system temperature of the instrument and the detected RFI occupancy, the setting of the Galaxy and the Sun overshadow the influence caused by true RFI sources, and this is the only structured variation over time that is apparent in the data. Therefore, RFI is not a factor for deciding whether to observe at day or night. Of course, there are other reasons to conduct low-frequency observations at night, especially because of the stronger effect of the ionosphere and the presence of the Sun during the day, which both make successful calibration more challenging.

We estimate the false-positives rate of the AOFlagger pipeline to be 0.5--0.6\%, based on the level of falsely detected samples in clean-appearing data. The resulting loss in sensitivity is therefore negligible. We have seen that during long observations, in which the system temperature changes due to the setting of the Galaxy and the Sun, time ranges with increased variance result in higher levels of false detections. Therefore, it would be a good practice to apply the correction method that was used for the LBA set: by (temporarily) dividing the samples by an accurate estimate of the standard deviation before flagging the data, the rate of false-positives will become constant for timesteps with a different sky temperature. This requires two runs of the flagger: one run to be able to estimate the variance on clean data, and one more to flag the data with the normalised standard deviation. This decreases the level of false-positives by about 0.5\% (a total detected rate of 1.77\% instead of 2.24\%) on LBA sets and will also decrease the number of false negatives in areas of low variance, but because of the smaller field of view of the HBA array, the improvement is less significant there. It is computationally twice as expensive, and is not necessary for short observations that do not show a significant change in sky temperature.

Up to now, interference detection was often performed manually and ad-hoc by the observer. Consequently, few statistics are available in the literature that describe the amount of data loss in cross-correlated data due to interference for a particular observatory and frequency range, but some studies have been performed. A systematic analysis of interference at the Mauritius Radio Telescope showed an average RFI occupancy of 10\% \citep{statistics-of-mauritius-rt}. In general, compared to data losses achieved with common RFI excision strategies, the loss in LOFAR data is low. This is especially surprising considering the fact that LOFAR is built in a populated area and operates at low frequency. Several reasons can be given for the small impact of RFI on LOFAR:
\begin{itemize}
 \item Many interfering sources contaminate a narrow frequency range or short duration. LOFAR's high time and frequency resolutions, of 1~s and 0.76~kHz respectively, minimise the amount of data loss caused by such interfering sources. Since the current loss of data is small, it seems unnecessary to go to even higher resolutions.
 \item LOFAR is the first telescope to use many novel post-correlation detection methods, such as the scale-invariant rank operator and the {\tt Sum\-Thresh\-old} techniques, which allow detection with high accuracy.
 \item LOFAR's hardware is designed to deal with the strong interfering sources that are found in its environment. The receiver units remain in linear state in the neighbourhood of such sources, and the strong band-pass filters spectrally localise the sources. Consequently, almost no interfering source will cause ramifications in bands that are adjacent to their transmitting frequency. The only exception is at very low frequencies, where we do see a very strong source saturate the ADCs when ionospheric conditions are bad. This source and its harmonics are successfully removed during flagging.
 \item Propagation models for Earth-bound signals show a strong dependence on the height of the receiver (e.g., \citet{hata-propagation-loss}). In contrast to dishes with feeds in the focal point, the receiving elements of LOFAR are close to the ground.
 \item LOFAR is remotely controlled, and the in situ cabins with electronics are shielded. We have found no post-correlation contamination that is caused by self-generated interference. This is in contrast with for example the WSRT, where the dishes close to the control room (which contains the correlator, but it is operated from elsewhere) are known to observe more interference. In the LOFAR auto-correlations, every now and then we do see some artefacts that suggest local interference, but these do not visibly contaminate cross-correlations. It might be that forming station beams before correlation helps reducing such RFI as well.
\end{itemize}

Given the low impact of RFI on LOFAR, we can conclude that the interference environment should not have an absolute weight in site selection of future (low-frequency) radio telescopes --- or its substations --- for example for the Square Kilometre Array. Instead, it should be carefully weighted against the non-negligible costs of logistics that are involved in building and maintaining a telescope in a remote area, and when dealing with low frequencies, against the quality of the ionosphere for performing radio astronomy.

In this article, we have not yet looked at the Gaussianity of the signal and the implications of the statistical distribution of RFI. Such statistical properties of RFI sources might have implications on long integrations, such as the LOFAR EoR project. We will deal with this in future work.

\bibliography{lofarradioenvironment}

\begin{thebibliography}{29}
\expandafter\ifx\csname natexlab\endcsname\relax\def\natexlab#1{#1}\fi

\bibitem[{Baan {et~al.}(2004)Baan, Fridman, \& Millenaar}]{wsrt-rfims}
Baan, W.~A., Fridman, P.~A., \& Millenaar, R.~P. 2004, {AJ}, 128, 933

\bibitem[{Barnbaum \& Bradley(1998)}]{adaptive-cancellation}
Barnbaum, C. \& Bradley, R.~F. 1998, AJ, 115, 2598

\bibitem[{{Bentum} {et~al.}(2008){Bentum}, {Boonstra}, {Millenaar}, \&
  {Gunst}}]{bentum-lofar-rfi-mitigation}
{Bentum}, M., {Boonstra}, A.-J., {Millenaar}, R., \& {Gunst}, A. 2008, in URSI
  General Assembly 2008 (Chicago: URSI)

\bibitem[{Bentum {et~al.}(2010)Bentum, Boonstra, \&
  Millenaar}]{assessment-of-LOFAR-RFI}
Bentum, M.~J., Boonstra, A.~J., \& Millenaar, R.~P. 2010, Proc. of RFI2010

\bibitem[{Boonstra(2005)}]{boonstra-dissertation}
Boonstra, A.~J. 2005, PhD thesis

\bibitem[{Bregman(2000)}]{bregman-lofar-concept-design}
Bregman, J.~D. 2000, in SPIE Proc. Astr. Telescopes and Instr., ed. H.~R.
  Butcher, Vol. 4015 (SPIE), 19--32

\bibitem[{Cappellen {et~al.}(2005)Cappellen, Bregman, \&
  Arts}]{dipole-array-sensitivity-cappellen}
Cappellen, W.~A., Bregman, J.~D., \& Arts, M.~J. 2005, Exp. Astr., 17, 101

\bibitem[{de~Bruyn {et~al.}(2011)de~Bruyn, Brentjens, Koopmans, Zaroubi,
  Labropoulos, \& Yatawatta}]{de-bruyn-eor-ursi-2011}
de~Bruyn, A.~G., Brentjens, M.~A., Koopmans, L. V.~E., {et~al.} 2011, in
  General Assembly and Scientific Symposium, 2011 XXXth URSI, 1--4

\bibitem[{de~Vos {et~al.}(2009)de~Vos, Gunst, \& Nijboer}]{lofar-system-design}
de~Vos, M., Gunst, A.~W., \& Nijboer, R. 2009, Proceedings of the IEEE, 97,
  1431

\bibitem[{Ellingson \& Hampson(2002)}]{ellingson-spatial-nulling-2002}
Ellingson, S.~W. \& Hampson, G.~A. 2002, {IEEE Trans. on Antennas \&
  Propagation}, 50, 25

\bibitem[{{Fl{\"o}er} {et~al.}(2010){Fl{\"o}er}, {Winkel}, \&
  {Kerp}}]{effelsberg-rfi-mitigation}
{Fl{\"o}er}, L., {Winkel}, B., \& {Kerp}, J. 2010, in Proc. of RFI2010

\bibitem[{Hata(1980)}]{hata-propagation-loss}
Hata, M. 1980, "IEEE Trans. on Vehicular Technology", VT-29, 317

\bibitem[{Kocz {et~al.}(2012)Kocz, Bailes, Barnes, Burke-Spolaor, \&
  Levin}]{spatial-filtering-parkes-multibeam-for-pulses}
Kocz, J., Bailes, M., Barnes, D., Burke-Spolaor, S., \& Levin, L. 2012, Monthly
  Notices of the Royal Astronomical Society, 420, 271

\bibitem[{Kocz {et~al.}(2010)Kocz, Briggs, \&
  Reynolds}]{spatial-filtering-parkes-multibeam}
Kocz, J., Briggs, F.~H., \& Reynolds, J. 2010, AJ, 140, 2086

\bibitem[{Leshem {et~al.}(2000)Leshem, van~der Veen, \&
  Boonstra}]{multichannel-rfi-mitigation}
Leshem, A., van~der Veen, A.-J., \& Boonstra, A.-J. 2000, ApJS, 131, 355

\bibitem[{{Niamsuwan} {et~al.}(2005){Niamsuwan}, {Johnson}, \&
  {Ellingson}}]{pulse-blanking}
{Niamsuwan}, N., {Johnson}, J.~T., \& {Ellingson}, S.~W. 2005, {Radio Science},
  40

\bibitem[{Offringa {et~al.}(2010b)Offringa, {de Bruyn}, Biehl, \&
  Zaroubi}]{LOFAR-RFI-pipeline}
Offringa, A.~R., {de Bruyn}, A.~G., Biehl, M., \& Zaroubi, S. 2010b, in Proc.
  of RFI2010, Astron (PoS)

\bibitem[{Offringa {et~al.}(2010a)Offringa, {de Bruyn}, Biehl, Zaroubi,
  Bernardi, \& Pandey}]{post-correlation-rfi-classification}
Offringa, A.~R., {de Bruyn}, A.~G., Biehl, M., {et~al.} 2010a, MNRAS, 405, 155

\bibitem[{Offringa {et~al.}(2012{\natexlab{a}})Offringa, de~Bruyn, \&
  Zaroubi}]{post-correlation-filtering}
Offringa, A.~R., de~Bruyn, A.~G., \& Zaroubi, S. 2012{\natexlab{a}}, MNRAS,
  422, 563

\bibitem[{Offringa {et~al.}(2012{\natexlab{b}})Offringa, van~de Gronde, \&
  Roerdink}]{scale-invariant-rank-operator}
Offringa, A.~R., van~de Gronde, J.~J., \& Roerdink, J. B. T.~M.
  2012{\natexlab{b}}, A\&A, 539

\bibitem[{Pandey \& Shankar(2005)}]{statistics-of-mauritius-rt}
Pandey, V.~N. \& Shankar, N.~U. 2005, in Proc. of XXVIIIth URSI General
  Assembly

\bibitem[{Romein(2008)}]{bandpass-correction-lofar-romein}
Romein, J.~W. 2008, Bandpass correction in {LOFAR}, Tech. rep., ASTRON

\bibitem[{Romein {et~al.}(2011)Romein, Mol, van Nieuwpoort, \&
  Broekema}]{blue-gene-romein}
Romein, J.~W., Mol, J.~D., van Nieuwpoort, R.~V., \& Broekema, P.~C. 2011, in
  URSI General Assembly, 2011 (IEEE), 1--4

\bibitem[{Ryabov {et~al.}(2004)Ryabov, Zarka, \&
  Ryabov}]{exoplanet-detection-with-rfi}
Ryabov, V., Zarka, P., \& Ryabov, B. 2004, Planetary and Space Science, 52,
  1479

\bibitem[{Smolders \& Hampson(2002)}]{hampson-spatial-nulling-2002}
Smolders, B. \& Hampson, G. 2002, {IEEE Antennas \& Propagation magazine}, 44,
  13

\bibitem[{Stappers {et~al.}(2011)Stappers, Hessels, Alexov,
  {et~al.}}]{lofar-pulsars}
Stappers, B.~W., Hessels, J. W.~T., Alexov, A., {et~al.} 2011, A\&A, 530, A80

\bibitem[{Weber {et~al.}(1997)Weber, Faye, Biraud, \&
  Dansou}]{chi-square-time-blanking-weber}
Weber, R., Faye, C., Biraud, F., \& Dansou, J. 1997, A\&AS, 126, 161

\bibitem[{Wijnholds {et~al.}(2005)Wijnholds, Bregman, \&
  Boonstra}]{sky-noise-limited-wijnholds}
Wijnholds, S.~J., Bregman, J.~D., \& Boonstra, A.-J. 2005, Exp. Astr., 17, 35

\bibitem[{Wijnholds \& van Cappellen(2011)}]{lofar-antenna-performance}
Wijnholds, S.~J. \& van Cappellen, W.~A. 2011, IEEE Trans. on Antennas and
  Propagation, 59, 1981

\end{thebibliography}
\bibliographystyle{aa}

\end{document}